\documentclass[aps,showpacs,floatfix,twocolumn]{revtex4-1}
\usepackage{amsmath}
\usepackage{graphicx}
\usepackage{epsfig}
\usepackage{lscape}
\usepackage{ulem}
\def\beq{\begin{equation}}
\def\eeq{\end{equation}}
\def\beqa{\begin{eqnarray}}
\def\eeqa{\end{eqnarray}}
\def\ban{\begin{eqnarray*}}
\def\ean{\end{eqnarray*}}
\def\bi{\begin{itemize}}
\def\ei{\end{itemize}}

\begin{document}

\title{Dynamical properties of nuclear and stellar matter and the symmetry energy}
\author{Helena Pais, Alexandre Santos,  Luc\'ilia Brito and Constan{\c c}a Provid\^encia}
\affiliation{Centro de F\'{\i}sica Computacional, Department of Physics\\
University of Coimbra, 3004-516 Coimbra, Portugal}
\begin{abstract}
The effects of density dependence of the symmetry energy on the
collective modes and  dynamical instabilities of cold and warm
nuclear and stellar matter are studied in the framework of
relativistic mean-field hadron models. The existence of the
collective isovector and possibly an isoscalar collective mode
above saturation density is discussed. It is shown that soft
equations of state do not allow for a high density isoscalar
collective mode, however, if the symmetry energy is hard enough an
isovector mode will not disappear at high densities. The
crust-core transition density and pressure are obtained as a
function of temperature for $\beta$-equilibrium matter with and
without neutrino trapping. An estimation of the size of the
clusters formed in the non-homogeneous phase as well as the
corresponding growth rates and distillation effect is made. It is
shown that cluster sizes increase with temperature, that the
distillation effect close to the inner edge of the crust-core
transition is very sensitive to the symmetry energy, and that,
 within a dynamical instability calculation, the pasta phase exists
in warm compact stars up to 10 - 12 MeV.
\end{abstract}


\maketitle
\section{Introduction}
Understanding the properties of
isospin rich
nuclear matter is one of the present goals
of both nuclear physics and astrophysics.
 A major scientific effort is being
carried out at an international level to study experimentally
the properties of asymmetric nuclear systems
and probe the behavior of the symmetry energy close to and above
saturation density \cite{Li-08}.  Astrophysical
observations of compact objects are also a window into both
the bulk and the microscopic properties of nuclear matter
at extreme isospin asymmetries \cite{Steiner-05}. Measurements coming from both laboratory and
astrophysical observations are expected to put constraints on the acceptable properties of the
equation of state (EOS) of asymmetric nuclear matter \cite{Lattimer-07,Baran-05}.

In the last years an important effort has been done to determine the
 density dependence of the symmetry energy of asymmetric nuclear matter (see the reviews
\cite{Baran-05,Steiner-05,Li-08} and references there in).
Correlations between different quantities in bulk matter and
finite nuclei have been established. Examples are the correlation between the slope of the pressure
of neutron matter at $\rho=0.1$ fm$^{-3}$ and the neutron skin thickness of $^{208}$Pb
\cite{BrownB-00,*Typel-01}, or the correlation between the crust-core transition density and the  neutron
skin thickness of $^{208}$Pb \cite{Horowitz-01}. Both Skyrme forces and relativistic mean-field (RMF) models have
been used to describe these correlations.
 Recently, it was shown that the parameters
characterizing   the microscopic Brueckner-Hartree-Fock (BHF)
equation of state of isospin asymmetric nuclear matter fall within
the trends predicted by several Skyrme and relativistic effective
models \cite{Vidana-09}. These models are compatible with recent constraints coming
from heavy ion collisions \cite{Li-08}, giant monopole resonances \cite{Garg-07}, or isobaric
analog states \cite{Danielewicz-09}. { In \cite{Vidana-09} the correlation between the slope $L$ and the curvature $K_{\rm sym}$ of the symmetry energy 
with the neutron skin thickness and the crust-core transition density in compacts stars was
analyzed  using the equation of state of the different models and  a thermodynamic approach to
determine  the transition density.
 Many other works  have studied the instability of asymmetric  matter using a thermodynamical approach  \cite{Miller-95,*Li-97,*Li-02}.
 
Instead of using a thermodynamic  approach, in \cite{Pethick-95a}  the transition density  was estimated in  a  local equilibrium approximation by determining the density for which  matter becomes unstable to  small density fluctuations. 
 It has been shown in \cite{Oyamatsu-07,Avancini-08,Avancini-09,Xu-09-ASJ}
 that the thermodynamical method gives a good estimation  of the transition density for cold
$\beta$-equilibrium matter, although a bit too large. In \cite{Avancini-08,Avancini-09}  a comparison was done between
the transition density obtained from the Vlasov formalism, a semi-classical limit of the description of the  collisionless regime,  and a Thomas-Fermi calculation of
the pasta phase. It was shown  that both results almost coincide:  the Vlasov
formalism predicts a transition density  only slightly smaller then the Thomas-Fermi one.  
Moreover, Ducoin {\it et al.} \cite{Ducoin-08-NPA} 
have shown  that  the transition density obtained within the Vlasov formalism does not differ from the one obtained by \cite{Pethick-95a} using a local equilibrium approximation.
}

{
The main goal of this work is to study, in the framework of relativistic models,  the effect of the symmetry
energy density dependence on some dynamical properties of
asymmetric nuclear matter, namely the  dynamical
instabilities at sub-saturation densities at zero and finite temperature and the  isoscalar and isovector
collective modes in asymmetric nuclear matter at zero temperature.
}

{It is important to  know the properties of the crust  to learn about neutron stars. For instance,  seismic events in the neutron star crust are sensitive to the shear modulus of the crust, which is itself  sensitive to the composition of the neutron star crust, namely the proton and neutron contents of the clusters in the inner crust \cite{Strohmayer-91}. Also, the cooling and evolution of the neutron star crust depends on its size and transport properties, both related with the composition of the star \cite{Lattimer-94,*Potekhin-97,*BrownE-00}.}

 As shown in \cite{Pethick-95,Link-99} the pressure at the
inner boundary of the crust defines the mass and moment of
inertia of the crust. This establishes  a relation between the EOS and compact stars observables.
Until now the transition pressure and density have only been studied at zero temperature but it is also of interest to 
  determine how do these quantities depend on the temperature. This will be one of the aims of
the present work.
 The transition density and pressure will be determined from the crossing of
the $\beta$-equilibrium EOS of stellar matter and the dynamical spinodal. Warm stellar matter
may contain trapped neutrinos \cite{Prakash-97}. We will consider the conditions  expected in the inner crust and in
the core of a neutron star, namely of
isospin asymmetry,  density and temperature. {This same problem was investigated  very recently
within a momentum-dependent effective interaction for nucleons \cite{Xu-10}.

}

Collective modes in asymmetric nuclear matter have already been studied in previous works
\cite{Haensel-78,Matera-94,Colonna-98,Greco-03,Avancini-05,Providencia-06} both within
non-relativistic and relativistic nuclear models.  These modes corresponding to isospin or density waves in infinitely
extended nuclear matter can be identified as the  counterpart of giant resonances in the atomic
nucleus \cite{Haensel-78}. They may be  of particular importance for the formation of shock waves
in nuclear matter \cite{Scheid-74,Haensel-78} and they will affect the thermal properties of neutron stars, since the neutrino mean free paths are affected by the propagation of collective modes in the medium \cite{Reddy-99}.

The authors of Ref.\cite{Greco-03} have predicted the existence of a neutron wave at
reasonably large densities, for $\rho \sim 3\rho_0$, where $\rho_0$ is the saturation density.
They have also shown that the inclusion 
of  the scalar
isovector virtual $\delta$($a_0$(980)) field may strongly reduce the restoring force for the isovector collective modes of the system.
 In a similar way they predict that, if the incompressibility is
decreasing with the increase of the density, the exotic high density isoscalar collective modes
could disappear. This could be a  signature of the softening of the equation of state (EOS) at
large densities \cite{Greco-03}.

 In the present work we will  study the collective modes of nuclear matter 
within the Vlasov equation for relativistic nuclear models delevoped in
\cite{Nielsen-91,Nielsen-93} for symmetric nuclear matter and applied later to asymmetric
nuclear matter \cite{Avancini-05,ProvidenciaC-06,Providencia-06}. It is  a semi-classical
approach which can be used to determine the eigenmodes of nuclear asymmetric matter and
corresponds to the $\hbar\to 0$ limit  of the time dependent Hartree-Fock. 
 The relativistic Vlasov equation for the non-linear Walecka model (NLWM)  was first applied
to  the study of heavy-ion collisions by Ko and collaborators \cite{Ko-87,*Ko-88} and to  describe the
collective modes of symmetric nuclear matter in \cite{Nielsen-91}.
In this last work, it was shown that in the low momentum transfer  limit this formalism gives results  similar to the ones obtained  using the mean-field approximation for the  ground-state and calculating the meson propagators in the one-loop approximation \cite{Lim-89}.

In order to discuss the dynamical response of asymmetric nuclear matter we will
 consider  a
set of NLWM  that differ in the density dependence of the
symmetry energy and the incompressibility.
 The following models will be considered: NL3
\cite{Lalazissis-97} with a quite large symmetry energy and
incompressibility at saturation and which was fitted in order to
reproduce the ground state properties of both stable and unstable
nuclei, TM1 \cite{Sumiyoshi-95} which also reproduces the ground
state   properties of both stable and unstable nuclei and provides
an equation of state of nuclear matter similar to the one obtained
in the RBHF (Relativistic Brueckner Hartree-Fock) theory, softer
than NL3 at high densities, FSU \cite{Todd-Rutel-05} which was
accurately calibrated to simultaneously describe the GMR in
$^{90}$Zr and $^{208}$Pb, and the IVGDR in $^{208}$Pb and still
reproduce ground-state observables of stable and unstable nuclei.
In TM1 a quartic self-interaction $\omega$-meson term was
included,  which softens the EOS.  This term is also present in
the FSU parametrization together with a  mixed isoscalar-isovector
coupling which modifies the density dependence of the symmetry
energy. Finally, we also consider NL3$\omega\rho$ and TM1$\omega\rho$, the NL3 and TM1 parametrizations with a
mixed isoscalar-isovector coupling which we will vary in order to
change the density dependence of the symmetry energy
\cite{Horowitz-01}.
The EOS obtained within NL3 is too hard and well above the limits that  collective flow data \cite{Danielewicz-02}
 impose to the EOS of symmetric matter.
Also,  the symmetry energy slope of NL3 at saturation is 118 MeV slightly above  the
limit obtained with experimental constraints from  isospin diffusion $63<L<113$ MeV, and well
above other constraints \cite{Centelles-09}. Nevertheless, we have chosen this model
because it has been calibrated to describe the groundstate of a wide number of stable and
unstable nuclei and because it is still frequently used.  
This model is taken as an example of a model with an hard EOS both in the isoscalar and isovector channel.

We will consider both asymmetric nuclear matter,  neutral  neutron-proton-electron ($npe$)
neutrino-free matter in $\beta$-equilibrium at zero temperature and $npe$  matter with trapped
neutrinos in $\beta$-equilibrium for a lepton fraction $Y_l=0.4$, both at zero and finite temperature.

In Sec. \ref{II} we review the formalism used,  in Sec. \ref{III}
we present and discuss the results obtained and finally, in Sec.
\ref{V} we draw some conclusions.

\section{The Vlasov equation formalism} \label{II}
We use the relativistic non-linear Walecka model (NLWM) in the mean-field approximation, within the Vlasov formalism to study nuclear collective modes of asymmetric nuclear matter and  $npe$ matter at finite temperature \cite{Nielsen-91,Nielsen-93}.

We consider a system of baryons, with mass $M$ interacting with
and through an isoscalar-scalar field $\phi$ with mass $m_s$, an
isoscalar-vector field $V^{\mu}$ with mass $m_v$ and an
isovector-vector field $\mathbf b^{\mu}$ with mass $m_\rho$. When
describing $npe$ matter we also include a system of electrons with
mass $m_e$. Protons and electrons interact through the
electromagnetic field $A^{\mu}$. The Lagrangian density reads:
$$
{\cal L}=\sum_{i=p,n} {\cal L}_i + {\cal L}_e + {\cal L}_{\sigma} + {\cal L}_{\omega}  + {\cal L}_{\rho} + {\cal L}_A
$$
where the nucleon Lagrangian reads
$$
{\cal L}_i=\bar \psi_i\left[\gamma_\mu i D^{\mu}-M^*\right]\psi_i \, ,
$$
with
$$
i D^{\mu}=i\partial^{\mu}-g_v V^{\mu}-
\frac{g_{\rho}}{2}  {\boldsymbol\tau} \cdot \mathbf{b}^\mu - e A^{\mu}
\frac{1+\tau_3}{2} \,,
$$
$$
M^*=M-g_s \phi  \, ,
$$
and the electron Lagrangian is given by
$$
{\cal L}_e=\bar \psi_e\left[\gamma_\mu\left(i\partial^{\mu} + e A^{\mu}\right)
-m_e\right]\psi_e .
$$
The isoscalar part is associated with the scalar sigma ($\sigma$)
field $\phi$, and the vector omega ($\omega$) field $V_{\mu}$,
whereas the isospin dependence comes from  the isovector-vector
rho ($\rho$) field $b_\mu^i$ (where $\mu$ stands for the four
dimensional space-time indices  and $i$ the three-dimensional
isospin direction index). The associated Lagrangians are:
\begin{eqnarray}
{\cal L}_\sigma&=&+\frac{1}{2}\left(\partial_{\mu}\phi\partial^{\mu}\phi
-m_s^2 \phi^2 - \frac{1}{3}\kappa \phi^3 -\frac{1}{12}\lambda\phi^4\right),\nonumber\\
{\cal L}_\omega&=&-\frac{1}{4}\Omega_{\mu\nu}\Omega^{\mu\nu}+\frac{1}{2}
m_v^2 V_{\mu}V^{\mu} + \frac{1}{4!}\xi g_v^4 (V_{\mu}V^{\mu})^2, \nonumber \\
{\cal L}_\rho&=&-\frac{1}{4}\mathbf B_{\mu\nu}\cdot\mathbf B^{\mu\nu}+\frac{1}{2}
m_\rho^2 \mathbf b_{\mu}\cdot \mathbf b^{\mu}, \nonumber\\
{\cal L}_A&=&-\frac{1}{4}F_{\mu\nu}F^{\mu\nu},
\end{eqnarray}
where
$\Omega_{\mu\nu}=\partial_{\mu}V_{\nu}-\partial_{\nu}V_{\mu} ,
\quad \mathbf B_{\mu\nu}=\partial_{\mu}\mathbf b_{\nu}-\partial_{\nu} \mathbf b_{\mu}
- g_\rho (\mathbf b_\mu \times \mathbf b_\nu)$ and $F_{\mu\nu}=\partial_{\mu}A_{\nu}-\partial_{\nu}A_{\mu}$.

The model comprises the following parameters:
three coupling constants $g_s$, $g_v$ and $g_{\rho}$ of the mesons
to the nucleons, the bare nucleon mass $M$, the electron mass
$m_e$, the masses of the mesons, the electromagnetic coupling
constant $e=\sqrt{4 \pi/137}$ and the self-interacting coupling
constants $\kappa$, $\lambda$ and $\xi$. In this Lagrangian
density, $\boldsymbol \tau$ is the isospin operator.

The NL3$\omega\rho$, TM1$\omega\rho$ and FSU parametrizations also include a
nonlinear $\omega\rho$ coupling term, which allows to change the
density dependence of the symmetry energy. This term is given by:
\begin{equation}
{\cal L}_{\omega \rho}=\Lambda_{v} g_v^2 g_\rho^2 \mathbf
b_{\mu}\cdot \mathbf b^{\mu}\, V_{\mu}V^{\mu}
\end{equation}
The $\omega$-meson self-interacting term, with the $\xi$ coupling
constant, is present only in the TM1, TM1$\omega\rho$ and FSU parametrizations.

We have used the  set of constants FSU \cite{Piekarewicz-09},
TM1 \cite{Sumiyoshi-95},  NL3
\cite{Lalazissis-97},   NL3$\omega\rho$ model
\cite{Horowitz-01}, with $\Lambda_{v}=0.01,\, 0.02,\,0.03$.
The same values of $\Lambda_{v}$ have been used in TM1$\omega\rho$.
 The saturation density that we refer
as $\rho_0$ is 0.148 fm$^{-3}$, except for TM1, which is
$\rho_0=0.145$ fm$^{-3}$.

\begin{table*}[t]
  \begin{tabular}{c c c c c c c c}
    \hline
    \hline
 Model&$\rho_{0}$&$E/A$&$K$&$E_{sym}$&$L$&$K_{sym}$&$K_{\tau}$   \\
    \hline
FSU&0.148&-16.302&227.895&32.537&60.395&-51.415&-275.570\\
TM1&0.145&-16.265&279.514&36.835&110.607&33.557&-517.390\\
TM1$\omega\rho$,$\Lambda_{v}=0.01$&0.145&-16.265&279.514&34.855&85.290&-79.145&-503.986\\
TM1$\omega\rho$,$\Lambda_{v}=0.02$&0.145&-16.265&279.514&33.235&67.654&-97.150&-434.142\\
TM1$\omega\rho$,$\Lambda_{v}=0.03$&0.145&-16.265&279.514&31.880&55.082&-72.164&-346.532\\
NL3&0.148&-16.240&269.937&37.344&118.320&100.525&-696.129\\
NL3$\omega\rho$,$\Lambda_{v}=0.01$&0.148&-16.240&269.937&34.929&87.638&-46.2956&-636.362\\
NL3$\omega\rho$,$\Lambda_{v}=0.02$&0.148&-16.240&269.937&33.121&68.147&-53.458&-512.296\\
NL3$\omega\rho$,$\Lambda_{v}=0.03$&0.148&-16.240&269.937&31.660&55.227&-8.069&-379.915\\
    \hline
    \hline
  \end{tabular}
\caption{ Nuclear matter properties at saturation density,
$\rho_0$. All quantities are in MeV, except for $\rho_0$, given in
fm$^{-3}$. }  \label{7}
\end{table*}

Table \ref{7} shows  nuclear matter properties at the
saturation density for these models: the binding energy per
nucleon $E/A$, the incompressibility coefficient $K$, the symmetry energy ${\mathcal{E}}_{{sym}}$,
the symmetry energy slope $L$, the symmetry energy curvature $K_{{sym}}$ and
$K_{\tau}=K_{sym}-6L-\frac{Q_0}{K}L$ (see Ref.\cite{Vidana-09}).

\begin{figure*}[t]
 \begin{tabular}{cc}
   \includegraphics[angle=0,height=0.3\linewidth]{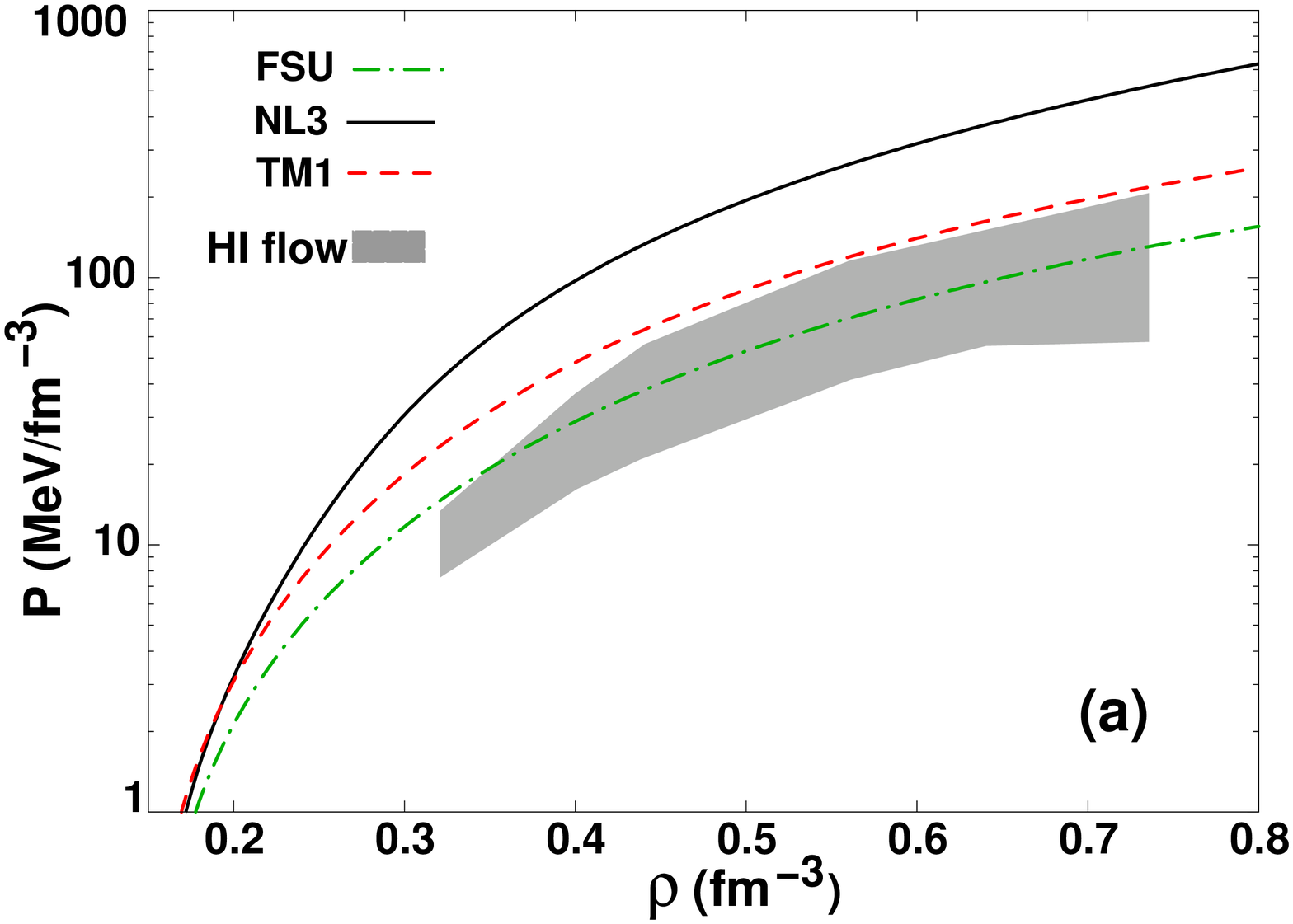}&
   \includegraphics[angle=0,height=0.3\linewidth]{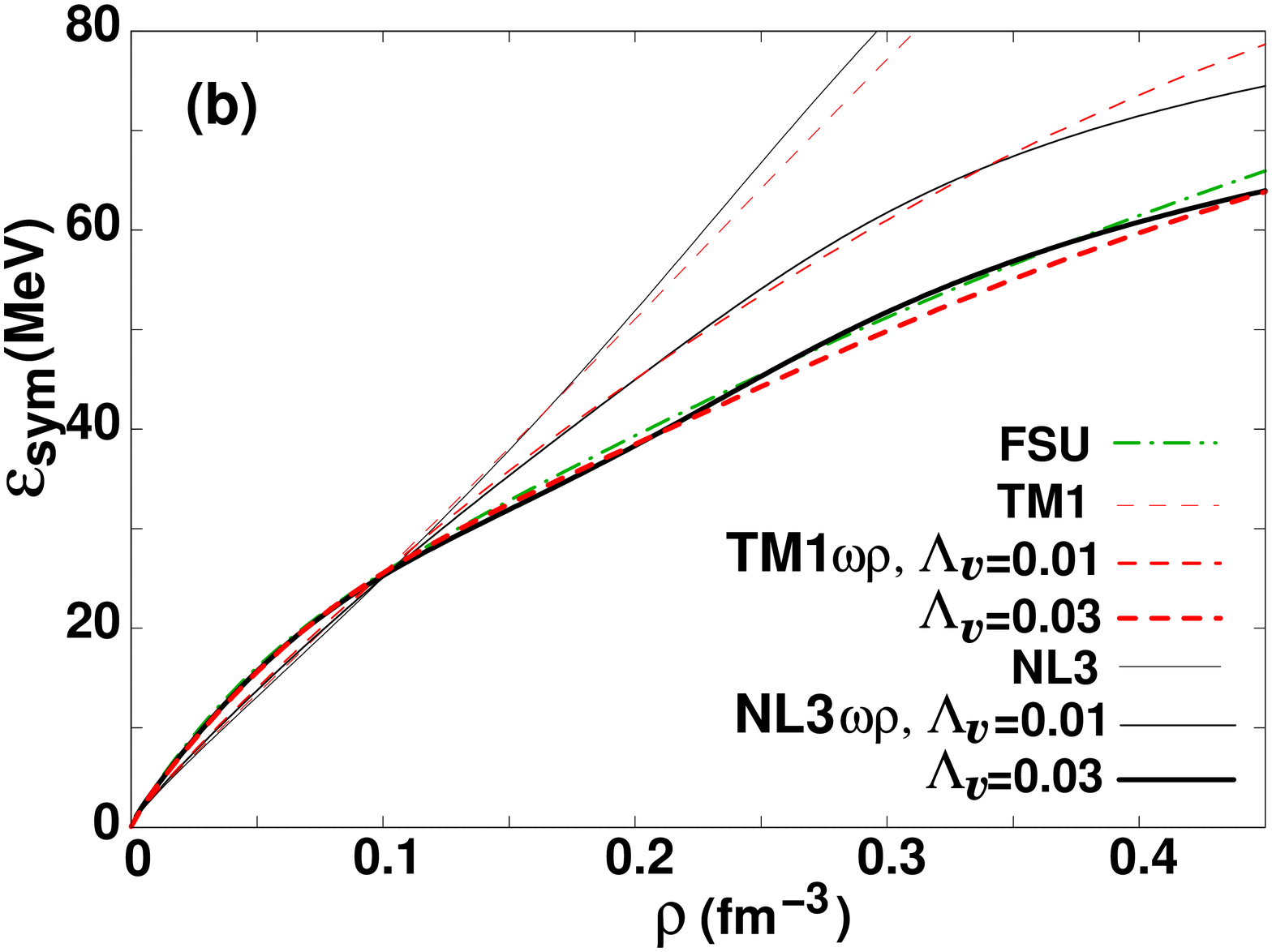}
 \end{tabular}
\caption{ (Color online) a) Pressure of symmetric nuclear matter
for NL3, TM1 and FSU and b) symmetry energy for NL3  and NL3$\omega\rho$, TM1 and
TM1$\omega\rho$, and FSU. We have considered 
$\Lambda_{v}=0.01$, 0.03 in  NL3$\omega\rho$ and TM1$\omega\rho$. For symmetric matter
the EOS of NL3$\omega\rho$ and TM1$\omega\rho$ coincide respectively with NL3 and TM1. In  a) we also
include the constraints obtained from the flow analysis in heavy
ion collisions (shaded region) \cite{Danielewicz-02} .
}%
\label{fig1}
\end{figure*}

We have chosen a set of models which have different
incompressibilities at large densities and different density
dependences of the symmetry energy. This is clearly seen in Fig.
\ref{fig1} where we have plotted the pressure of symmetric nuclear
matter, Fig. \ref{fig1}a), and  the symmetry energy for densities
below 0.45 fm$ ^{-3}$, Fig. \ref{fig1}b). In Fig. \ref{fig1}a) we
also include the constraints  obtained from the flow analysis in
heavy ion collisions \cite{Danielewicz-02}.  It is seen that NL3
and  NL3$\omega\rho$ models give a too hard EOS at large
densities, which is not compatible with the constraints from heavy
ion collisions. FSU is well inside the constraints and TM1 and TM1$\omega\rho$ are just
slightly above the upper limit. 

{ The non-linear $\omega\rho$ term will only affect the
symmetry energy, therefore, models that only differ by an extra  non-linear $\omega\rho$ term
will have the same EOS for symmetric matter.}
As we will show later, this
behavior will have a strong effect on the high energy collective
modes. With respect to the symmetry energy NL3 and TM1 have a very
hard behavior at high densities while FSU is softer. The
NL3$\omega\rho$ and TM1$\omega\rho$ models show an intermediate behavior and approach
FSU when the $\omega\rho$ coupling parameter
$\Lambda_v$ increases. Below $\rho=0.1$ fm$^{-3}$ the softness of the
symmetry energy is reversed with NL3 the softest and FSU the
hardest.

In the sequel we use the formalism developped in ref. \cite{Nielsen-93} where the collective modes
in hot and dense symmetric nuclear matter were determined within the Vlasov equation
based on the Walecka model \cite{Walecka-74}. We will use, whenever possible, the same notation.
We start from the  thermodynamical potential density for $npe$ matter
$$\Omega={\cal E}-\sum_{i=n,p,e} \mu_i \rho_i - T \varsigma$$
where $\cal E$ is the energy density
\begin{eqnarray}
{\cal E}&=&2\int\frac{ d^3p}{(2\pi)^3}\,\nonumber\\
&&\left[f({\bf r},{\bf p},t)_+ \,h({\bf r},{\bf
  p},t)_+ - f({\bf r},{\bf p},t)_- \,h({\bf r},{\bf
  p},t)_-\right]\nonumber\\
&+&\frac{1}{2}\left(\Pi_\phi^2\,+\,{\bf\nabla}\phi\cdot
{\bf\nabla}\phi\,+\,m_s^2\phi^2 +\frac{1}{3}\kappa \phi^3 +
\frac{1}{12}\lambda\phi^4\right)\nonumber\\
&+&\frac{1}{2}[\Pi_{V_i}^2\,-\,2\Pi_{V_i}
\partial_i V_0\,+\,{\bf\nabla} V_i\cdot{\bf\nabla} V_i\,\,\nonumber\\
&&-\partial_j V_i\partial_i V_j\,+\,m_v^2({\bf V}^2 - V_0^2)\,+\,\frac{1}{4}\xi g_v^4({\bf V}^2-V_0^2)^2] \nonumber\\
&+&\frac{1}{2}[\Pi_{b_i}^2\,-\,2\Pi_{b_i}
\partial_i b_0\,+\,{\bf\nabla} b_i\cdot{\bf\nabla} b_i\nonumber\\
&&-\partial_j b_i\partial_i b_j\,+\,m_{\rho}^2({\bf b}^2 - b_0^2)]\nonumber \\
&+&\frac{1}{2}\left(\Pi_{A_i}^2\,-\,2\Pi_{A_i}
\partial_i A_0\,+\,{\bf\nabla} A_i\cdot{\bf\nabla} A_i-\partial_j
A_i\partial_i A_j\right)\nonumber\\
&+&\Lambda_{v} g_v^2 g_\rho^2 ({\bf V}^2 - V_0^2)({\bf b}^2 - b_0^2),\nonumber\\
\end{eqnarray}
with       $\Pi_{\phi}$ ($\Pi_{V_i}$, $\Pi_{b_i}$, $\Pi_{A_i}$ ) the field canonically
conjugated to $\phi$ ($V_i$, $b_i$, $A_i$ ),
$\varsigma$ is the entropy density
\begin{eqnarray}
\varsigma&=&2\sum_{i=p,n,e}\int\frac{ d^3p}{(2\pi)^3}\,
\left[f_{i+} \, \mbox{ln}\left(f_{i+}\right)\right. \nonumber\\
&+& \left.\left(1-f_{i+}\right) \,
  \mbox{ln}\left(1-f_{i+} \right)+ f_{i-} \, \mbox{ln}\left(f_{i-}\right)\right. \nonumber\\
&+& \left.\left(1-f_{i-}\right) \,
  \mbox{ln}\left(1-f_{i-} \right)\right]
\end{eqnarray}
and $\mu_i$ and $\rho_i$ are the chemical potential and particle
density of particle type $i$.
We only consider the third
component in isospin space of the $\rho$-meson
and
we denote by
$$f({\bf r},{\bf p},t)_{\pm}=\mbox{diag}(f_{p \pm},\,f_{n \pm},\,f_{e \pm})$$
the  distribution function of particles (+) at position $\mathbf
r$, instant $t$ and momentum $\mathbf{p}$ and of antiparticles
($-$) at position $\mathbf r$, instant $t$ and momentum
$-\mathbf{p}$, and by
\begin{equation}
h_{\pm}=\mbox{diag}(h_{p\pm},h_{n\pm},h_{e\pm})
\end{equation}
the corresponding one-body Hamiltonian, where
\begin{equation}
h_{i\pm}=\pm \sqrt{({\bf p}-{\boldsymbol{\cal V}}_i)^2+M^{*2}}+ {\cal V}_{0i}.
\label{aga}
\end{equation}
For protons and neutrons, $i=p,n$, we have
\begin{eqnarray*}
{\cal V}_{0i}&=& g_v V_0  + \frac{g_\rho}{2}\tau_i b_0+ e A_0 \frac{1+\tau_{i}}{2} ,\\
{\boldsymbol{{\cal V}}}_{i}&=& g_v {\mathbf V} +
\frac{g_\rho}{2}\, \tau_i {\mathbf b}+ e\,  {\mathbf A}
\frac{1+\tau_{i}}{2},
\end{eqnarray*}
with $\tau_i=1$ (protons) or $-1$ (neutrons). For electrons,
$i=e$, we have
$$
{\cal V}_{0e}=-eA_0,\qquad {\boldsymbol {\cal V}}_e=-e {\bf A},\qquad M_e^*=m_e.
$$

The time evolution of the distribution functions is described by the
Vlasov equation
\begin{equation}
\frac{\partial f_{i \pm}}{\partial t} +\{f_{i \pm},h_{i \pm}\}=0, \qquad
\; i=p,\,n,\, e,
\label{vlasov1}
\end{equation}
where $\{,\}$ denote the Poisson brackets.
The Vlasov equation  expresses the
conservation of the number of particles in phase space and, therefore,
 at a given temperature, the dynamics described by the
Vlasov equation does not change the number of particles or antiparticles in
this semiclassical approach.

From Hamilton's equations we derive the equations describing the time
evolution of the fields $\phi$,  $V^\mu$, $A^\mu$, and the third component of
the $\rho$-field, $b_3^\mu=(b_0,\mathbf{b})$, which are given in Appendix \ref{eq_campos} (Eqs. (\ref{eqmphi})-(\ref{eqma})).
The state which minimizes the thermodynamical potential density of $npe$ matter
is characterized by the distribution functions
\begin{equation}
f_{0 i \pm}= \frac{1}{1+e^{(\epsilon_{0 i} \mp \nu_i)/T}}, \quad i=p,n,
\label{fpn}
\end{equation}
with
\begin{equation*}
\epsilon_{0 i}=\sqrt{p^2+{M^*}^2},  \quad \nu_i=\mu_i - g_v V_0  - \frac{g_\rho}{2}\, \tau_i b_0 - e\, A_0
\frac{1+\tau_i}{2}
\end{equation*}
and
\begin{equation}
f_{0 e \pm}= \frac{1}{1+e^{(\epsilon_{0 e} \mp \mu_e)/T}} \,\, ,
\label{fe}
\end{equation}
with
\begin{equation*}
\epsilon_{0e}=\sqrt{p^2+m_e^2}
\end{equation*}
and by the  constant mesonic fields, which obey the following equations:
$m_s^2\phi_0^{(0)} + \frac{\kappa}{2} \phi_0^{(0)\,2} +
\frac{\lambda}{6} \phi_0^{(0)\,3}=g_s\rho_s^{(0)},\,\,\,$
$m_v^2\,V_0^{(0)}+\frac{1}{6}\xi g_v^4 V_0^{(0)\,3}+2 \Lambda_{v}
g_v^2 g_\rho^2 V_0^{(0)} b_0^{(0)\,2}=g_v j_0^ {(0)},\,\,\,$
$m_{\rho}^2\,b_0^{(0)}+2 \Lambda_{v} g_v^2 g_\rho^2 V_0^{(0)\,2}
b_0^{(0)}=\frac{g_\rho}{2} j_{3,0}^{(0)},\,\,\,$
$V^{(0)}_i=b_i^{(0)}= A_0^{(0)}= A_i^{(0)}=0,\,\,$
where $\rho_s^{(0)}, \, \,  j_0^ {(0)},\, \, j_{3,0}^{(0)}$ are respectively the equilibrium
scalar density, nuclear density and isospin density.
Notice that for $T=0$ MeV the distribution functions, $f_{0i\pm}$, become $f_{0i+}=\theta(P_{Fi}^2-p^2)$ and the distribution functions for antiparticles vanish.

In order to evaluate the crust-core transition density we need to calculate the $\beta$-equilibrium  equation of state of $npe$ stellar matter. The pressure reads \cite{Avancini-06}
$$
P=\sum_{i=p,n} P_i + \sum_{l=e,\nu_e} P_l + P_{\sigma} + P_{\omega} + P_{\rho} + P_{\omega\rho}
$$
where
\begin{eqnarray}
\label{press}
P_i&=&\frac{1}{3\pi^2}\int dp \frac{p^4}{\sqrt{p^2+M_i^{*2}}}(f_{0i+}+f_{0i-}), \nonumber\\
P_l&=&\frac{1}{3\pi^2}\int dp \frac{p^4}{\sqrt{p^2+m_l^2}}(f_{0l+}+f_{0l-}), \nonumber\\
P_\sigma&=&-\frac{m_s^2}{2}\phi_0^{(0)\,2}-\frac{1}{6}k\phi_0^{(0)\,3}-\frac{1}{24}\lambda\phi_0^{(0)\,4},\nonumber\\
P_\omega&=&\frac{m_v^2}{2}V_0^{(0)\,2}+\frac{1}{24}\xi g_v^4 V_0^{(0)\,4}, \nonumber \\
P_\rho&=&\frac{m_\rho^2}{2}b_0^{(0)\,2}, \nonumber\\
P_{\omega\rho}&=&\Lambda_{v} g_v^2 g_\rho^2 V_0^{(0)\,2}
b_0^{(0)\,2}.
\end{eqnarray}
The chemical equilibrium conditions $\mu_p=\mu_n-\mu_e$ for neutrino free
matter, or $\mu_p=\mu_n-\mu_e+\mu_{\nu_e}$ for neutrino trapped matter, as
well as the electrical charge neutrality condition $\rho_e=\rho_p$ must be imposed.

Collective modes in the present approach correspond to small oscillations
around the equilibrium state. These small deviations are described by the
linearized equations of motion and collective modes are given as
solutions of those equations. To construct them, let us define:
\begin{eqnarray*}
f_{i \pm}&=&f_{0 i \pm}^{(0)} + \delta f_{i \pm}, \\ \nonumber
\phi_0\,&=&\,\phi_0^{(0)} + \delta\phi, \\ \nonumber
V_0\,&=&\, V_0^{(0)} + \delta V_0 , \quad V_i\,=\,\delta V_i, \\ \nonumber
b_0\,&=&\, b_0^{(0)} + \delta b_0, \quad b_i\,=\,\delta b_i, \\ \nonumber
A_0\,&=&\, \delta A_0, \quad A_i\,=\,\delta A_i.
\end{eqnarray*}

As in \cite{Nielsen-91,Nielsen-93,Avancini-05,Avancini-04}, we
express the fluctuations of the distribution functions in terms of the  generating
functions:
$$S_{\pm}({\mathbf r},{\mathbf p},t)=
\mbox{diag}\left(S_{p \pm},\,S_{n \pm},\,S_{e \pm}\right),$$
such that
$$\delta f_\pm=\{S_\pm,f_{0,\pm}\}=\{S_\pm,p^2\}\frac{df_{0 \pm}}{dp^2}.$$
The linearized Vlasov equations
for $\delta f_{i \pm}$,
$$\frac{d\delta f_{i \pm}}{d t}+ \{\delta f_{i \pm}, h_{0 i \pm}\}
 +\{f_{0 i \pm},\delta h_{i \pm} \}=0$$
 are equivalent to the following time-evolution equations:

\begin{eqnarray}
\label{eq:deltaf}
\frac{\partial S_{i \pm}}{\partial t} &+& \{S_{i \pm},h_{0i \pm}\} = \delta h_{i \pm}
= \mp \frac{g_s\, M_i^*}{\epsilon_{0 i}}\delta\phi \nonumber \\
&\mp& \frac{{\bf p} \cdot \delta \boldsymbol{\cal
V}_i}{\epsilon_{0 i}}  + \delta{\cal V}_{0i},\,\,i=p,n
\end{eqnarray}

\begin{equation}
  \label{eq:deltafe}
\frac{\partial S_{e \pm}}{\partial t} + \{S_{e \pm},h_{0e \pm}\} = \delta h_{e \pm}
= -e\left[ \delta{A}_{0} \mp \frac{{\bf p} \cdot \delta{\mathbf A}}{\epsilon_{0e}}\right],
\end{equation}
where
\begin{eqnarray*}
\delta{\cal V}_{0i}&=&g_v \delta V_0 + \tau_i \frac{g_{\rho}}{2}\,
\delta b_0 + e\, \frac{1+\tau_{i}}{2}\,\delta A_0 ,\\ \nonumber
\delta \boldsymbol{\cal V}_i&=& g_v \delta {\mathbf V} + \tau_i
 \frac{g_{\rho}}{2} \,\delta {\mathbf b}+ e\, \frac{1+\tau_{i}}{2}\,
\delta {\mathbf A},
\end{eqnarray*}
with $ h_{0 i \pm}\,= \pm \epsilon_{0 i} +{\cal V}^{(0)}_{0 i}$ and $h_{0 e \pm}=\pm \epsilon_{0 e}$.

Of particular interest on account of their physical relevance are
the longitudinal modes, with wave vector ${\bf k}$ and frequency $\omega$,
described by the ansatz
\begin{equation}
\left(\begin{array}{c}
S_{j \pm}({\bf r},{\bf p},t)  \\
\delta\phi  \\
\delta \zeta_0 \\ \delta \zeta_i
\end{array}  \right) =
\left(\begin{array}{c}
{\cal S}_{\omega\pm}^j (p,{\rm cos}\theta) \\
\delta\phi_\omega \\
\delta \zeta_\omega^0\\ \delta \zeta_\omega^i
\end{array} \right) {\rm e}^{i(\omega t - {\bf k}\cdot
{\bf r})} \;  ,
\label{ansatz}
\end{equation}
where $j=p,\, n,\, e$, $\zeta=V,\, b,\, A$ represent the
vector-meson fields and $\theta$ is the angle between ${\bf p}$
and ${\bf k}$. The wave vector of the excitation mode,  ${\bf k}$,
is identified with the momentum transferred to the system through
the process which gives rise to the excitation.

For the longitudinal modes,
we get $\delta V_\omega^x = \delta V_\omega^y =0\,$, $\delta b_\omega^x =
\delta b_\omega^y = 0\,$ and $\delta A_\omega^x = \delta A_\omega^y
=0\,$.
Calling $\delta V_\omega^z = \delta V_\omega$, $\delta b_\omega^z = \delta
b_\omega$ and $\delta A_\omega^z = \delta A_\omega$, we will  have for the fields
$ \delta {\cal V}_{i,z}= \delta {\cal V}_\omega^i{\rm e}^{i(\omega t - {\bf k}\cdot
{\bf r})}$ and
$\delta {\cal V}_{0i}= \delta {\cal V}_\omega^{0i}{\rm e}^{i(\omega t - {\bf k}\cdot
{\bf r})}.$
 Replacing the ansatz (\ref{ansatz})
in Eqs. (\ref{eq:deltaf}) and  (\ref{eq:deltafe}) we get
\begin{eqnarray}
i\left(\omega \mp \frac{k p \cos \theta}{\epsilon_{0 i}}\right )
{\cal S}_{\omega \pm}^i &=& \mp \frac{g_s\,M_i^*}{\epsilon_{0 i}}\delta\phi_\omega \nonumber \\
&+&\left[1 \mp \frac{\omega}{k}\frac{p \cos \theta}{\epsilon_{0
i}} \right]\delta {\cal V}_{ \omega}^{0i}  . \label{Si}\\
i\left(\omega \mp \frac{k p \cos \theta}{\epsilon_{0
e}}\right){\cal S}_{\omega \pm}^e &= & -e\left[1 \mp
\frac{\omega}{k}\frac{p \cos \theta}{\epsilon_{0 e}}
\right]\delta{A}_\omega^0, \qquad  \label{Se} \
\end{eqnarray}
In Appendix \ref{eq_flut} we present the equations of motion for
the field fluctuations (Eqs. (\ref{fi})-(\ref{contA})). The
solutions of these equations and Eqs. (\ref{Si}), (\ref{Se}) form
a complete set of eigenmodes that may be used to construct a
general solution for an arbitrary longitudinal perturbation.
Substituting the set of equations (\ref{fi})-(\ref{A}) into
(\ref{Si}) and (\ref{Se}) we get a set of equations for the
unknowns ${\cal S}_{\omega \pm}^i$, which lead to the following
matrix equation
\begin{equation}
\left(\begin{array}{ccccc}
a_{11}&a_{12}&a_{13}&a_{14}&a_{15}\\
a_{21}&a_{22}&a_{23}&a_{24}&0\\
a_{31}&a_{32}&a_{33}&a_{34}&a_{35}\\
a_{41}&a_{42}&a_{43}&a_{44}&0\\
0&0&a_{53}&0&a_{55}\\
\end{array}\right)
\left(\begin{array}{c}
\rho^S_{\omega p} \\
\rho^S_{\omega n} \\
\rho_{\omega p} \\
\rho_{\omega n} \\
\rho_{\omega e}
\end{array}\right)=0. \label{matriz}
\end{equation}
The coefficients  $a_{ij}$ are defined in Appendix \ref{rel_disp},
and the amplitudes $\rho^S_{\omega i}$ (Eq.(\ref{rhosi})) and
$\rho_{\omega i}$ (Eq.(\ref{rhoi})) are functions of the
quantities ${\cal S}_{\omega \pm}^i$.
 The dispersion relation is obtained from the determinant of the matrix of the coefficients, namely,
\begin{equation}
\mathrm{Det} (a_{ij})=0 \label{det}.
\end{equation}
The eigenmodes  $\omega$ of the system, with sound velocity $v_s=\omega/k$,  are solutions of the Eq. (\ref{det}). At
subsaturation densities, there are unstable modes identified by imaginary
frequencies. For these modes we define the growth rate $\Gamma=-i\omega$.
The region in ($\rho_p,\rho_n$) space for a given wave vector $\mathbf k$ and
temperature $T$,  limited by the surface $\omega=0$,
defines the dynamical spinodal surface.  
 In the  $k=0$ MeV limit we recover the thermodynamic
spinodal which is defined by the surface in the ($\rho_p,\, \rho_n,\,
T$) space for which the curvature matrix of the free energy density is
zero, i.e., has a zero eigenvalue. This relation has been
discussed in \cite{ProvidenciaC-06}.
\begin{figure*}[t]
 \begin{tabular}{c}
   \includegraphics[angle=0,width=1.\linewidth]{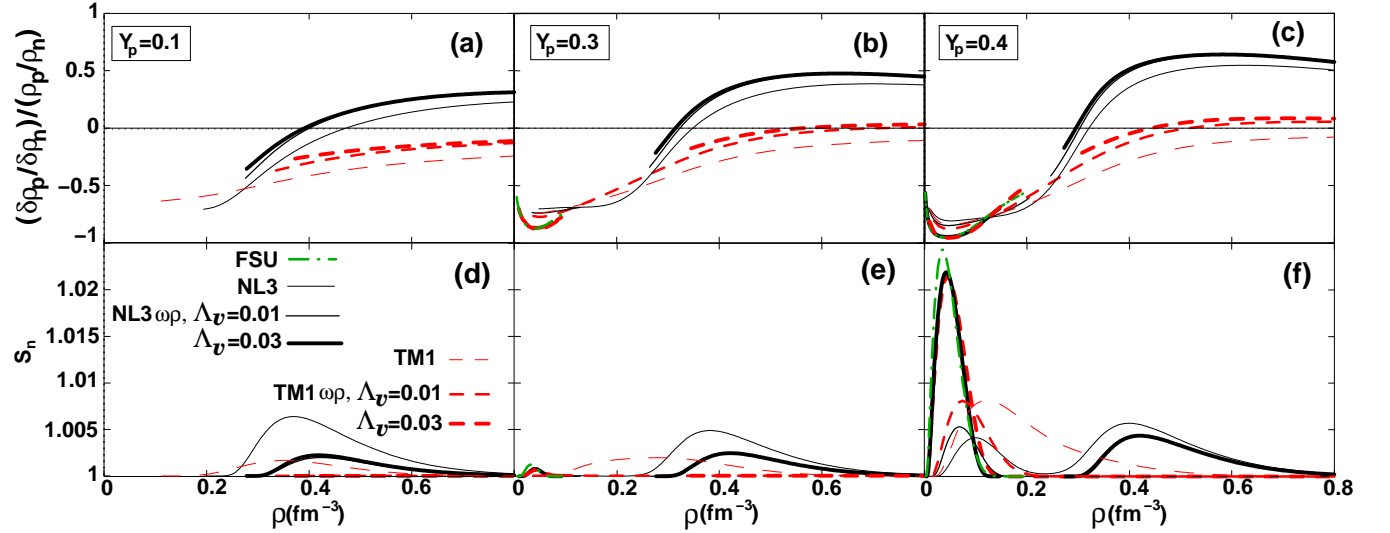}\\
 \end{tabular}
\caption{
(Color online)  The ratio of the proton to neutron density fluctuations
[Eq. (\ref{amp1})] divided by the ratio of the proton to neutron densities [(a) $y_p=0.1$, (b)
$y_p=0.3$ and (c) $y_p=0.4$] and
   the sound velocity $v_s$ normalized to the neutron Fermi velocity, $s_n=v_s/v_{Fn}$, [(d) $y_p=0.1$, (e)
$y_p=0.3$ and (f)  $y_p=0.4$]
 for NL3 and NL3$\omega\rho$,
TM1 and TM1$\omega\rho$, and FSU. $\Lambda_v=0.01$ and 0.03 have been used in NL3$\omega\rho$
and TM1$\omega\rho$. The curves have been obtained for  
 $T=0$ MeV and $k=200$ MeV.
}%
\label{fig2}
\end{figure*}

\section{Numerical results and discussions} \label{III}

In the present section we discuss the effect of the symmetry energy and incompressibility  on
some of the  dynamical properties of nuclear and stellar matter. We  calculate the
collective modes at densities above and below saturation density  and analyze their isoscalar or
isovector character.  We will next determine the dynamical  spinodal surfaces, for different
temperatures and wave vectors,
characterized by a zero frequency solution of the dispersion relation
(\ref{det}). They will allow us to predict under which conditions a non-homogeneous phase at
the crust of a proto-neutron star may exist.
 We will finally determine for warm $\beta$-equilibrium matter with trapped neutrinos
the unstable modes with the largest growth rate, which define the
mode that drives the system to a non-homogeneous phase and,
therefore, gives an estimation of the size of the clusters formed
in the phase transition. For these modes the magnitude of the
distillation effect will be calculated. For $\beta$-equilibrium
warm stellar matter with trapped neutrinos we will consider a
fixed lepton fraction $Y_l=0.4$, which corresponds to the maximum
expected value \cite{Prakash-97,Menezes-04a}.

\subsection{Collective modes}
We have calculated at zero temperature the normal modes of asymmetric nuclear matter as a function of the nuclear matter density $\rho$ and for different proton fractions
$y_p=\rho_p/\rho$.
Fig. \ref{fig2} shows in the top row  
 the ratio of the proton to
neutron density fluctuations, Eq. (\ref{amp1}), and in the bottom row  the
collective mode sound velocity, $s_n=v_s/v_{Fn}$, where $v_s$ is the sound velocity and $v_{Fn}$ is the Fermi velocity of neutrons. 
Modes with  a sound velocity below the neutron Fermi velocity 
are strongly damped and are not represented. 
\begin{figure}
   \begin{tabular}{c}
   \includegraphics[angle=0,width=7.cm]{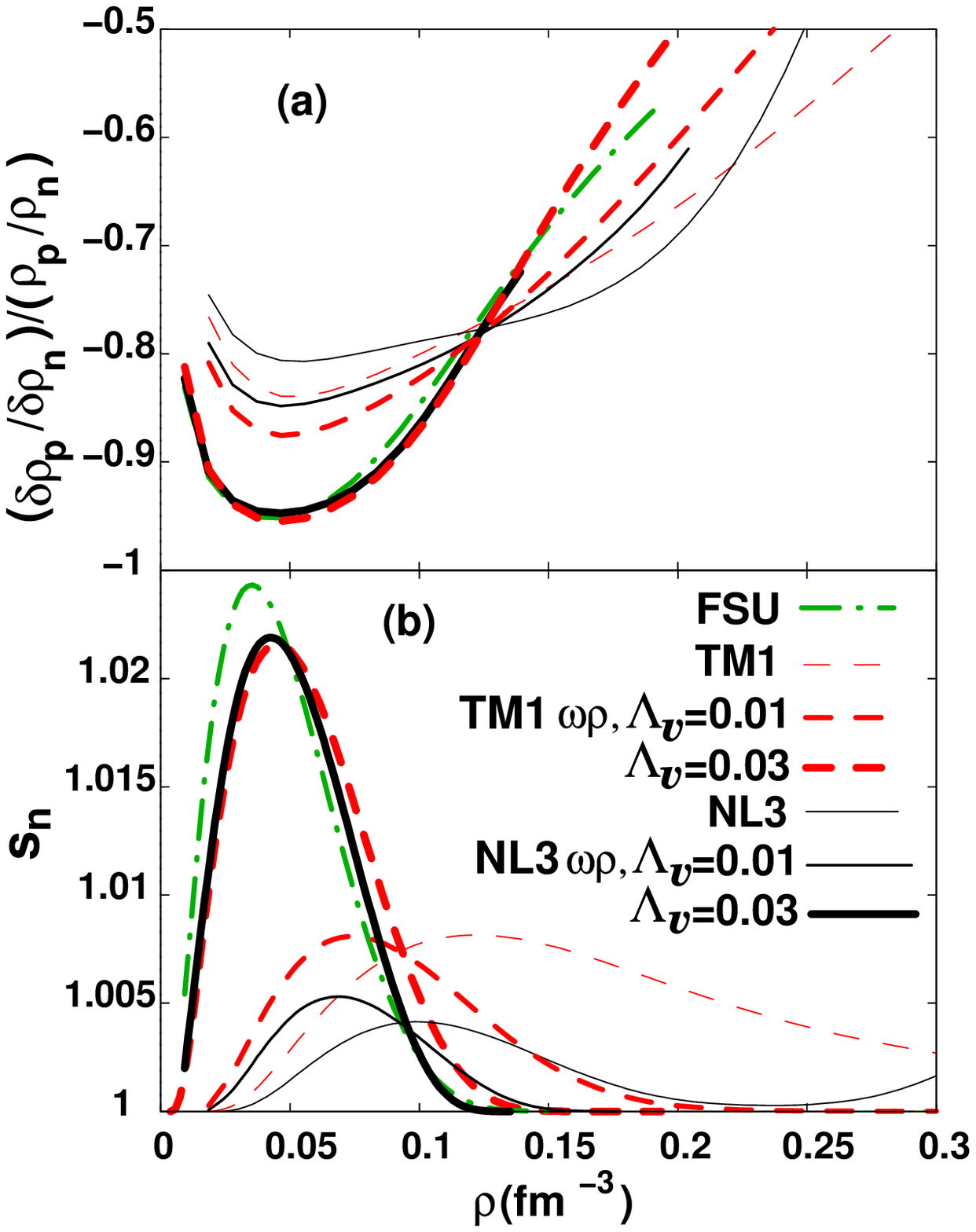}
   \end{tabular}
\caption{(Color online) Isovector-like mode below 2$\rho_0$ for $y_p=0.4$ and $k=200$ MeV:
a) the proton to neutron density fluctuation ratio normalized by the proton to neutron density
ratio  and b)   the sound velocity $v_s$ normalized to the neutron Fermi velocity, $s_n=v_s/v_{Fn}$.}
\label{fig2b}
\end{figure}

Several conclusions can
be drawn. Below $\rho=0.2$ fm$^{-3}$ and small isospin asymmetry
(large proton fractions) all models show an isovector like mode
which disappears for $y_p\lesssim 0.25$. In Fig. \ref{fig2b} the isovector response of the
system for $y_p=0.4$  is blown up for a better analysis of the results.
 By isovector
(isoscalar) like we mean that the ratio of the proton to neutron
density fluctuations is negative (positive), however, we stress that for asymmetric matter both
isoscalar and isovector components are mixed.

\begin{figure}
   \begin{tabular}{c}
   \includegraphics[angle=0,width=7.cm]{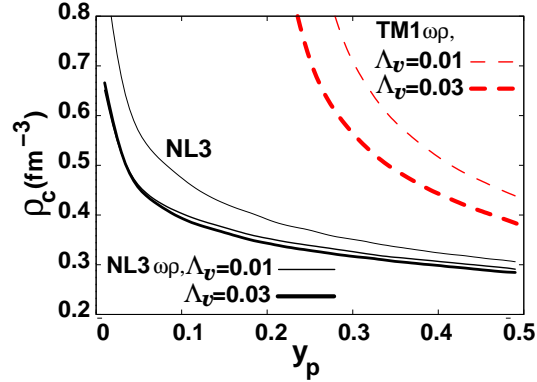}
   \end{tabular}
\caption{(Color online) The crossing density $\rho_c$ defining the onset of isoscalar character in
the collective modes in $np$ matter versus the proton fraction, for
k=200 MeV and the models  NL3,  NL3$\omega\rho$  and  TM1$\omega\rho$  with $\Lambda_{v}=0.01,\,
0.03$.} \label{fig3}
\end{figure}

 For NL3 and
NL3$\omega\rho$ the collective mode changes from an isovector like
to an isoscalar like mode at a density above twice the saturation
density, which we will designate by crossing density $\rho_c$. The
crossing density decreases with  an increase of the proton
fraction. This is clearly seen in Fig. \ref{fig3} where the
crossing density is plotted for NL3 and two parametrizations
of NL3$\omega\rho$ and TM1$\omega\rho$ ($\Lambda_v=0.01$ and 0.03). The crossing density is
sensitive to the high density dependence of the symmetry energy
and, therefore, the larger the crossing density the larger the
difference between NL3 and NL3$\omega\rho$.  
 We point out that NL3
and NL3$\omega\rho$ only differ on the isovector channel and
therefore, the differences among these models reflect the
isovector behavior. TM1 has, like NL3, a very stiff symmetry
energy, but, unlike NL3, has a quite soft EOS for symmetric matter
at large densities due to the inclusion of quartic $\omega$-terms.
As a result the collective mode exists at large densities but with
an isovector character. Including the non-linear $\omega\rho$ terms in TM1
changes this situation: for a strong enough coupling the isovector collective mode will change into an isoscalar mode, however, at much larger densities than NL3 and NL3$\omega\rho$ models, see Fig. \ref{fig3}.

 At very low densities the sound velocity of
the modes decreases as the asymmetry increases and, as seen in 
Fig.\ref{fig2}a) for $y_p=0.1$ there are no undamped modes  for any of the models. 
One immediate conclusion is the sensitivity of the normal modes to the density dependence of both the isoscalar and isovector EOS.
We will explain the behavior of normal modes with the density below.

\begin{figure}
   \begin{tabular}{c}
   \includegraphics[angle=0,width=7.cm]{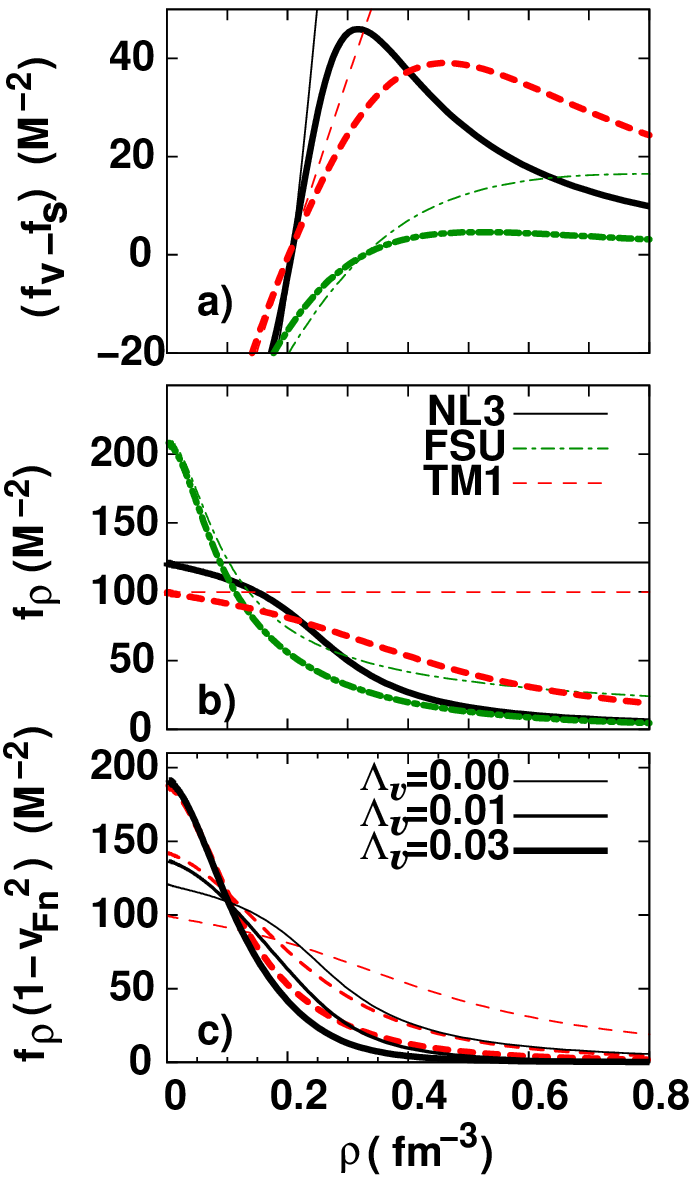}
   \end{tabular}
\caption{(Color online) The effective coupling parameters $f_i$ and
$f_i\, (1-v_F^2)$  in units of inverse nucleon mass
square ($M^{-2}$): a)  $f_v-f_s$  (thin lines)  and $(f_v-f_s) (1-v_F^2)$ (thick lines) for NL3, TM1 and FSU;
 b) $f_\rho$  (thin lines)  and $f_\rho\, (1-v_F^2)$ (thick lines) for NL3, TM1 and FSU; 
c)   $f_\rho\, (1-v_F^2)$ for NL3$\omega\rho$ and TM1$\omega\rho$ with $\Lambda_v=0,\, 0.01,\, 0.03$.}
\label{fig3a}
\end{figure}

The possible existence of isovector and isoscalar  modes was
discussed in \cite{Greco-03}. The authors of that paper have shown
that the existence of  modes would depend not only on the
saturation properties but also on the high density dependence of
the models. In particular, the isovector response of the system
could be significantly reduced at high densities due to the inclusion
of the $\delta$-meson. In a similar way the isoscalar mode depends
on the isoscalar response of the system and may not exist if the
EOS is too soft.

{The properties of the collective modes of nuclear matter described by the models
introduced in the first two sections are determined by the interaction factors $f_i$
\begin{equation}
 f_i=\left(\frac{g_i}{m_{i,eff}}\right)^2, \qquad i=s,\, v,\, \rho,
\end{equation} 
where the effective meson masses $m_{i,eff}$ are given by
\begin{eqnarray}
m^2_{s,eff}&=&m_s^2+\kappa\phi_0+\frac{\lambda}{2}\phi_0^2
+g_s^2\left(\frac{\partial\rho_s}{\partial M^*}\right)_0, 
\label{mseff}\\
m_{v, eff}^2 &=& m_{v}^2+ \frac{\xi}{2} g^4_v V_0^2 + 2\Lambda_v
g_v^2 g_\rho^2 b_0^2,
\label{mveff2}\\
m_{\rho,eff}^2&= &m_{\rho}^2+ 2 \Lambda_v g_v^2 g_\rho^2  V_0^2.
\label{mreff}
\end{eqnarray}

At $T=0$ MeV, the dispersion relation (\ref{det}) may be reduced to a simple analytical form 
for symmetric matter and in the low momentum limit \cite{Greco-03}. If we make the
approximation $v_s\approx v_{F}=k_{F}/\epsilon_F$  we get for the isoscalar mode
\beq
1+3\frac{\epsilon_F}{k_F^2} \, \rho
\left[(f_v-f_s) \, (1-v_F^2)+f_v f_s \,
v_F^2\frac{M^*}{\epsilon_F^2} \rho_s\right]\phi_s =0
\label{isoscalar}
\eeq
and for the isovector mode
\beq
1+\frac{3\epsilon_F}{k_F^2}\rho\, f_\rho (1-v_F^2) \phi_s=0,
\label{isovector}
\eeq
where $\phi_s$ is the Lindhard function $\phi_s=1-s/2 \, \ln |(s+1)/(s-1)|$, $s=v_s/v_F$.
}

We notice that  the second term of Eq. (\ref{isoscalar})
is dominated by the first part propotional to $(f_v -f_s)$, while  the second term
 of Eq. (\ref{isovector}) strongly depends on $f_\rho$. Both these factors are multiplied by
$1-v_F^2=(M^*/\epsilon_F)^2$ which approaches fast  zero when the density increases. In
Fig. \ref{fig3a} we represent these quantities:   $(f_v -f_s)$ and  $(f_v
-f_s)(1-v_F^2)$ in Fig. \ref{fig3a}a) and  $f_\rho$ and $f_\rho(1-v_F^2)$ in
Fig. \ref{fig3a}b) for NL3, TM1 and FSU, and $f_\rho(1-v_F^2)$ for TM1$\omega\rho$ and
NL3$\omega\rho$ in Fig. \ref{fig3a}c). There will be a isoscalar mode only if the factor that multiplies the
Lindhard function is positive. This only happens when the repulsion described by the
$\omega$-meson overcomes the $\sigma$-meson attraction, which occurs for $\rho\gtrsim 0.2$ fm$^{-3}$
for NL3 and TM1 and  for $\rho\gtrsim 0.3$ fm$^{-3}$ for FSU. However,  the increase of the
effective mass of the $\omega$-meson  in TM1 and FSU due to the quartic $\omega$-term in the
Lagrangian density [see Eq. (\ref{mveff2})], shows that this factor
is never so large as in NL3. The last term in  Eq. (\ref{mveff2}) is due to the $\omega\rho$
nonlinear coupling and also affects the effective $\omega$-meson mass for asymmetric matter.


The non-linear $\omega\rho$ term in NL3$\omega\rho$,  TM1$\omega\rho$ and FSU 
softens  the isovector
channel at large densities due to
an effective $\rho$-meson mass which increases with density.
As a consequence, the symmetry energy 
$$\varepsilon_{{sym}}=\frac{k_F^2}{6 \epsilon_F}+ \frac{1}{4} f_\rho \rho, $$ 
and the response of the
system to isovector excitations (\ref{isovector}),  which depend on 
$f_\rho$, will be significantly softened
at large densities.
Moreover, since the $g_\rho$ parameter is fixed so that at $\rho=0.1$ fm$^{-3}$ the symmetry energy
is the same for the models with and without the non-linear $\omega\rho$ term, the larger the coupling parameter $\Lambda_v$ the larger must be $g_\rho$. A larger
$g_\rho$ enhances the value of $f_\rho$ at small densities, and, therefore, also the response of the system to
isovector perturbations as seen in Fig. \ref{fig2b}: the models with the larger sound velocity
and the density fluctuation ratio close to -1 are the ones with the larger  $f_\rho$.

Comparing  $f_\rho$ with $f_\rho(1-v_F^2)$
[Fig. \ref{fig3a}b)], it is clear that the factor $1-v_F^2$ makes the second term of
(\ref{isovector}) go faster to zero at large densities. TM1 has the largest values of  $f_\rho(1-v_F^2)$ at large densities which may
explain its different behavior at these densities, namely the presence of an isovector like
mode.

We conclude that the main role of including the non-linear $\omega\rho$ term in NL3 and TM1 is to reduce the
value of $f_\rho$ at large densities, making the isovector character of the modes weaker in
this density range, and to increase the  $f_\rho$ at low densities  giving
rise to stronger isovector modes below $\rho\approx 0.2$ fm$^{-3}$. 

 We stress that the above discussion refers to
symmetric matter at zero temperature. In asymmetric matter isoscalar and isovector modes are
coupled and we expect to find the behavior of both pure isoscalar and pure isovector modes in all modes. A stronger  $f_\rho$ ($f_v -f_s$)  will give a stronger isovector (isoscalar) character to the mode. 

\begin{figure}[!]
 \begin{tabular}{c}
   \includegraphics[angle=0,width=7.cm]{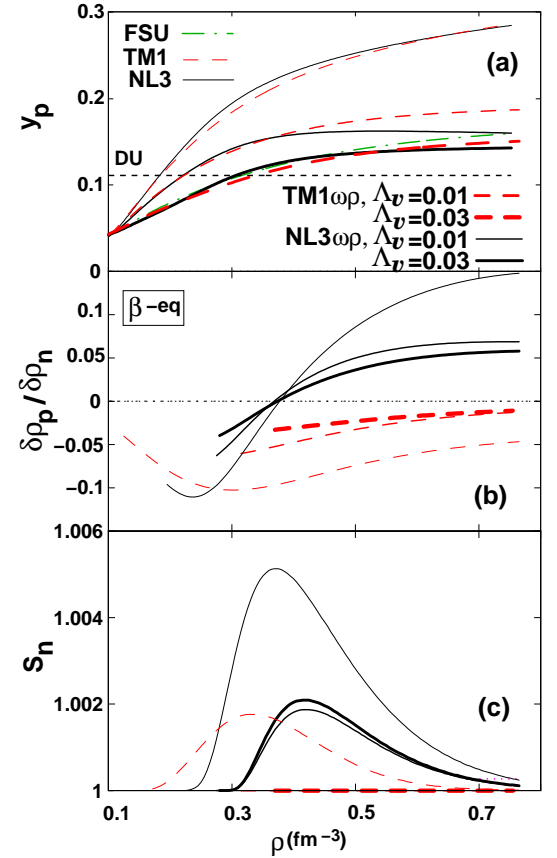}
 \end{tabular}
\caption{(Color online)  $\beta$-equilibrium stellar matter: (a) proton fraction  (b) the ratio of the proton-neutron density
fluctuations (Eq. (\ref{amp1})) and (c) the corresponding neutron sound velocities $s_n$ for
NL3,  FSU, TM1, and  NL3$\omega\rho$  and  TM1$\omega\rho$ with $\Lambda_{v}=0.01$ and
0.03. The results are for  $T=0$ MeV and $k=200\,$ MeV.}%
\label{fig4}
\end{figure}
In Fig. \ref{fig4}, we present the proton fractions 
(a) the proton-neutron density fluctuations (b) and the corresponding 
sound velocities of the collective modes $s_n=v_s/v_{Fn}$ (c) for $\beta$-equilibrium 
cold stellar matter.
For the FSU parametrization  we get no modes and for the parametrizations TM1 and TM1$\omega\rho$
only isovector-like modes are present. The other models change
character from isovector to isoscalar at the crossing  density
$\rho_c$. The crossing density in $\beta$-equilibrium matter  is
almost the same for all NL3 like  models 
 although, it was
seen in Fig. \ref{fig2} that the crossing density depends on the
isovector properties of the model. This is due to the fact that
the models have  different proton fractions for the same density
and there are two opposite effects that define the crossing
density: the larger the proton fraction the smaller the crossing
density and the larger the symmetry energy at saturation the
larger the crossing density. So the
combination of these two effects leads to a similar crossing
density for NL3 and  NL3$\omega\rho$.

\subsection{Subsaturation dynamical instabilities}
\begin{figure*}
   \begin{tabular}{c}
   \includegraphics[angle=0,width=16.cm]{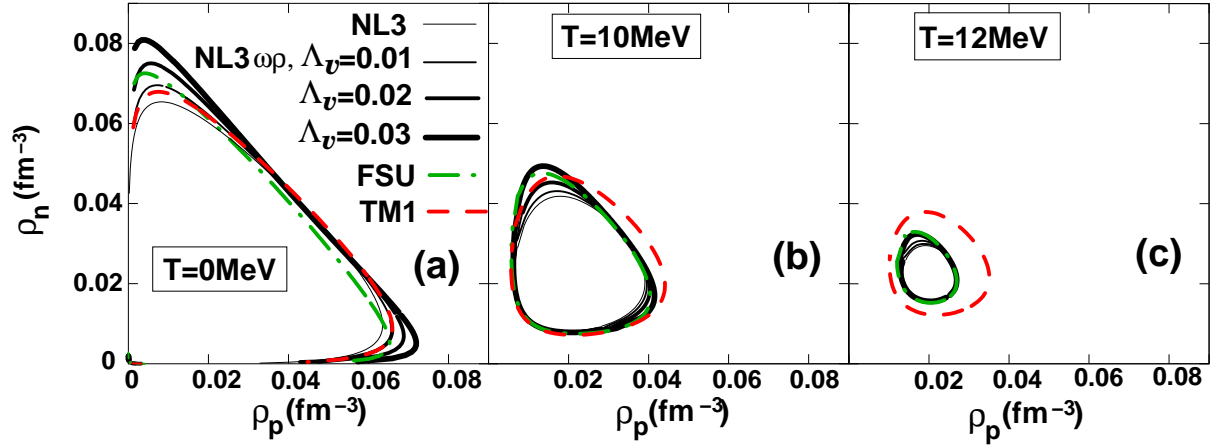}
   \end{tabular}
\caption{(Color online) Spinodals (Eq.(\ref{det}), with
$\omega=0$) for $k=75$ MeV for NL3, FSU, TM1 and NL3$\omega\rho$
with $\Lambda_{v}=0.01$,  0.02 and  0.03 at a) $T=0$, b) $T=10$
and  c) $T=12$ MeV.} \label{fig5}
\end{figure*}

We next analyze the properties of warm nuclear matter at
subsaturation densities, namely the range of densities for which
the system is unstable to density fluctuations. The surface on the
$T,\, \rho$, and $y_p$ space for which the sound velocity is zero
defines the dynamical spinodal. The extension of the spinodal
region depends on the wavelength of the perturbations: the system
is stable to small wavelength perturbation, of the order of the range
of the nuclear force or smaller and to large wavelengths due to
the contribution of the Coulomb interaction which blows up at
large wavelengths. Fig. \ref{fig5} shows the approximate
envelope of  the spinodal regions for different temperatures and
for NL3, FSU, TM1 and NL3$\omega\rho$ with
$\Lambda_{v}=0.01,0.02,0.03$. 
We will not consider TM1 with non-linear $\omega\rho$ because at subsaturation
 densities TM1 and NL3 behave in similar ways. 
With the increase of temperature,
the spinodal region shrinks to a region that is quite isospin
symmetric. TM1 has the highest critical temperature. For cold
matter the largest difference between the models occurs for very
isospin asymmetric, rich neutron matter. The extension of the
spinodal reflects both the value of the symmetry energy and its
density dependence. A larger symmetry energy and a larger slope of
the symmetry energy at saturation defines a smaller spinodal. Therefore, it
is expected that the extension of the non-homogeneous
phase in the inner crust of a compact star, the pasta phase, will
be sensitive to the symmetry energy.

The transition densities at the crust-core transition for
$\beta$-equilibrium warm stellar matter with trapped neutrinos
($Y_l=0.4$) are shown in Table
\ref{tab_press} for several temperatures. These densities are obtained from the crossing of
the EOS of $\beta$-equilibrium matter with the dynamical spinodal.
 The transition densities decrease with temperature due to
the decrease of the spinodal region. At $T=$0 MeV, there is no clear dependence of the transition densities on the slope $L$.
{However, if we consider the set of models that has the same isoscalar description, NL3 and
NL3$\omega\rho$, a clear tendency is seen:  $\rho_t$ decreases when $L$
increases. This is due to the corresponding decrease of the transition proton fraction  (for $T=$0 MeV $y_p=0.312,\,
0.314,\, 0.318,\, 0.322$, respectively, for NL3, NL3$\omega\rho$ with $\Lambda_v=0.01,\, 0.02,1, 0.03$).
A larger $L$ corresponds to a smaller symmetry energy at subsaturation densities and allows a larger asymmetry.

In Table \ref{tab_press} we also show the transition densities for
$\beta$-equilibrium neutrino-free stellar matter at $T=0$ MeV. For neutrino-free matter, the transition
densities are smaller  than the ones with trapped neutrinos because neutrino free matter has a much smaller proton
fraction.

 The pressure has no monotonic behavior with temperature: below $T\sim
5$ MeV it increases with temperature and only above $T\sim 5-10$ MeV it reduces with 
temperature due to a large reduction of the transition density. At lower
temperatures the small reduction of the transition density with respect to the $T=0$ MeV case
and a larger lepton contribution to the pressure explains the increase of the transition
pressure. Similar results have been obtained in \cite{Xu-10}.

 The effect of the symmetry energy slope $L$ on the transition pressure of matter
with trapped neutrinos is only clear for the set of models with the same isoscalar EOS, namely NL3 and NL3$\omega\rho$. For these models
the transition pressure increases with $L$ at $T=$0 MeV but does not show a systematic behavior at
$T=$5 and decreases with $L$ at $T=$ 10 MeV.
At $T=$0 MeV, the smaller isospin asymmetry
for the smaller $L$ corresponds to a  smaller pressure. This
tendency inverts as the temperature increases due to the larger lepton contribution to the pressure for the larger densities.

For neutrino-free matter at $T=$0 MeV it is seen that taking the whole set of models the transition pressure does not show a clear dependence with density, in agreement with the findings of \cite{Ducoin-10} with a much larger set of models. However, if we limit ourselves to the set NL3 and NL3$\omega\rho$ with the same isoscalar dependence the pressure decreases with density.
 }

\begin{table*}
\begin{tabular}{c c c c c c c c c c c cccccccccccc}
\hline
\hline
           &  $T$(MeV)  &      &      &  $\rho_{t}$    && (fm$^{-3}$)&  & &      &  &       &     $P_t$&  & &(MeV/fm$^3$) &    &  & & & \\
           &      &      &      &      &   & &  &      &    &   &   &    & && & &  & & &\\
\cline{3-9}\cline{11-19}
           &     &      &      &      & & & NL3$\omega\rho$   & &    &     &       &    &   & & & NL3$\omega\rho$& & &\\
           &     &      &      &      & & &  $\Lambda_{v}$   & &    &     &       &    &   & & &  $\Lambda_{v}$ & & &\\
           &     & FSU  & TM1  & NL3  & & $0.01$ & $0.02$ & $0.03$ & & FSU  & TM1   &  NL3  & && 0.01 & 0.02  & 0.03 & \\
\cline{3-5}\cline{7-9}\cline{11-13} \cline{16-19}
$Y_\nu$=0  &   0 &0.0745&0.0602&0.0538& & 0.0619 & 0.0741 & 0.0855 & &0.388 & 0.324 & 0.236 & && 0.357& 0.500 & 0.535 &\\
           &     &      &      &      & &        &   &    &   &      & & &&      &       & \\
$Y_l$=0.4  &   0 &0.0814&0.0829&0.0816& & 0.0828 &  0.0840 & 0.0850& &0.665 & 0.844 & 0.865 & && 0.838& 0.802 & 0.732 & \\

           &   5 &0.0789&0.0803&0.0781& & 0.0795 &  0.0809 & 0.0821& & 1.021 & 1.064 & 1.087 & && 1.074& 1.086 & 1.090 &\\

           &  10 &0.0652&0.0671&0.0605& & 0.0621 &  0.0642 & 0.0667 & &1.059 & 1.097 & 0.971 & && 1.022& 1.061 & 1.102 &\\

           &  12 &0.0475&0.0545& --   & &   --   &    --   & --     & &0.842 & 0.969 &  --   & && --   &  --   & --    &\\

\hline
\hline

\end{tabular}
\caption{The transition  densities and the pressures for NL3, TM1,
FSU and NL3$\omega\rho$ with $\Lambda_{v}=0.01,0.02,0.03$ at the
crust-core transition for $\beta$-equilibrium neutrino-free
stellar matter ($Y_\nu$=0) at $T=0$ MeV and for
$\beta$-equilibrium stellar matter with trapped neutrinos
($Y_l=0.4$) for several temperatures. } \label{tab_press}
\end{table*}
\begin{figure}[!htb]
\begin{tabular}{c}
\includegraphics[angle=0,width=7.cm]{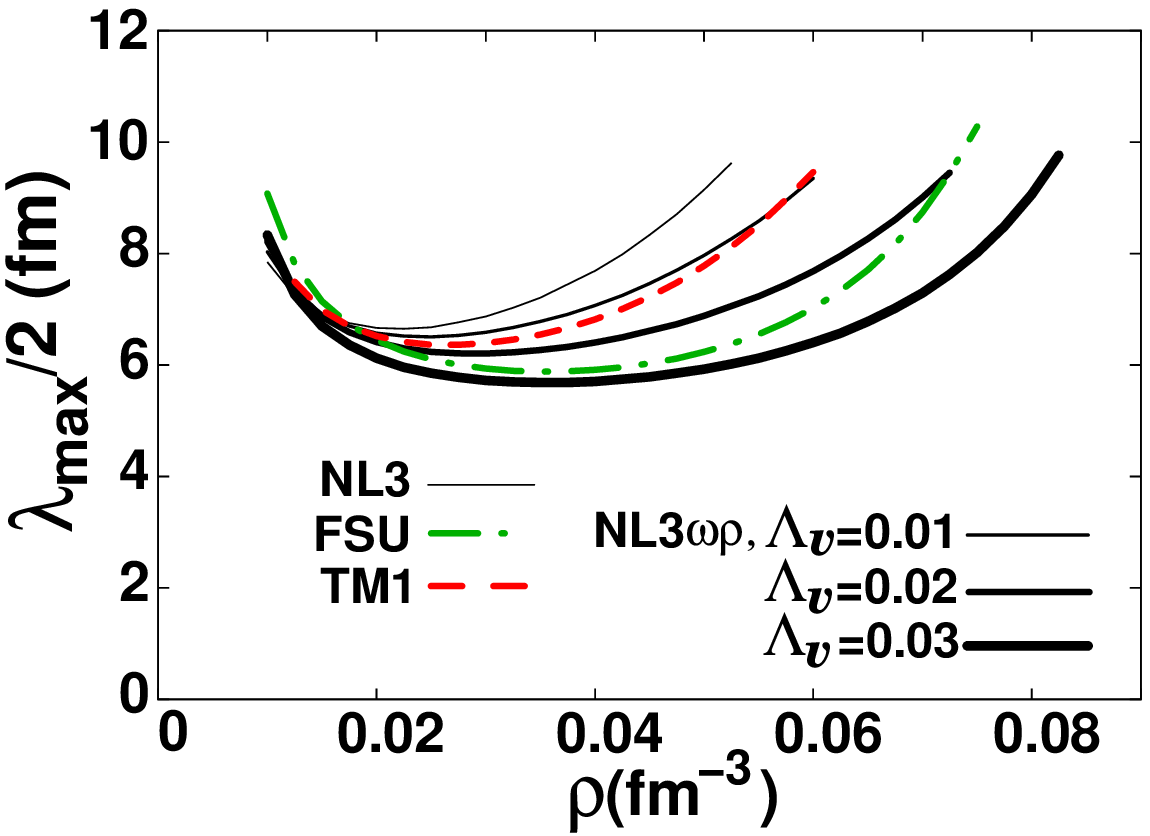}
\end{tabular}
\caption{(Color online) $\beta$-equilibrium matter ($Y_{\nu}=0$): estimated sizes of the
clusters at $T=0$ for NL3, FSU, TM1 (light brown, and NL3$\omega\rho$ with $\Lambda_{v}=0.01$, 0.02, 0.03.}%
\label{fig6}
\end{figure}
\begin{figure*}[!htb]
 \begin{tabular}{l}
   \begin{tabular}{c}
   \includegraphics[angle=0,width=16.cm]{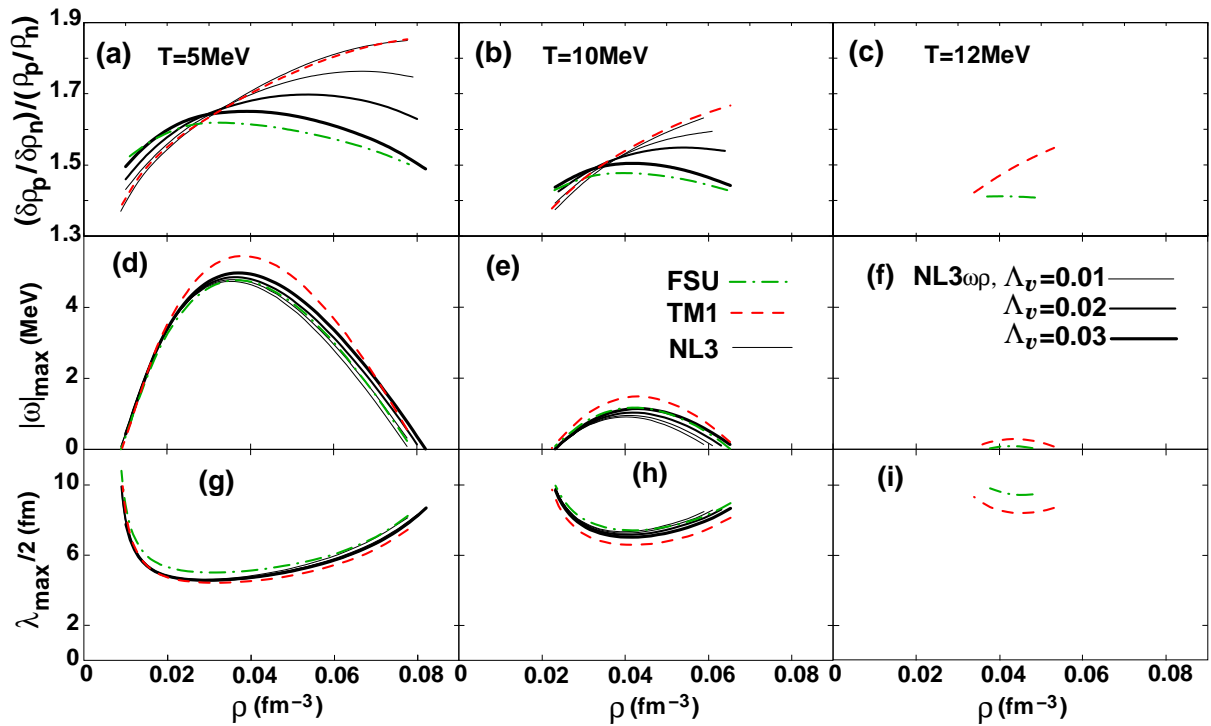}
   \end{tabular}
 \end{tabular}
\caption{(Color online)  $\beta$-equilibrium neutrino-trapped
matter ($Y_l=0.4)$: the normalized ratio of the proton over the neutron
density fluctuations for the most unstable mode  (largest $|\omega|$) at (a) $T=5$ , (b) $T=10$
and (c) $T=12$ MeV; the
corresponding growth rates at (d) $T=5$ , (e) $T=10$  and (f) $T=12$ MeV
and  estimated sizes of the clusters at (g) $T=5$, (h) $T=10$  and (i) $T=12$ MeV
for NL3, FSU, TM1
and NL3$\omega\rho$ with $\Lambda_{v}=0.01$, 0.02 and 0.03.}
\label{fig7}
\end{figure*}

The unstable modes of the system are calculated by replacing  the
frequency $\omega$  by  $i\Gamma$, where $\Gamma$ defines the
exponential growth rate of the instabilities. We take the mode
with the largest growth rate as the most unstable amongst all and,
therefore, the one that drives the system to the formation of
instabilities. Its half-wavelength defines the most probable size
of the clusters (liquid) formed in the mixed (liquid-gas) phase
\cite{Lucilia-06}. We would like to pointed out that the present
method of estimating the cluster size takes into account both the
Coulomb interaction,  which has been included in the formalism,
and  surface effects, which  are due to the finite range of the
nuclear force. Within a relativistic mean-field approach to
nuclei and finite clusters surface properties are determined by
the exchange of mesons with a finite mass. The estimation done
with the present method agrees with the results obtained within a
Thomas Fermi calculation of the pasta phase \cite{Avancini-08}.

In Fig. \ref{fig6} we show the cluster size estimates for matter in
$\beta$-equilibrium at zero temperature. Large differences for different models can be
seen  for the larger densities. This is due to the fact that cold  $\beta$-equilibrium
matter has quite small proton fractions, for which larger differences arise between the
models. It is also interesting to notice that close to the crust-core transition the clusters
get larger and all models predict a size of the order of 10 fm. There is also a close
correlation between the extension of the non-homogeneous phase determined by the density at
which each curve ceases to exist, and the slope of the symmetry energy as noticed by many
authors \cite{Xu-08,Vidana-09}. While below $\rho=0.02$ fm$^{-3}$ all the models predict the
same size of clusters, above this density there are differences between the models which are
reflected in the fact that all predict similar sizes at the transition density, although this
density differs from
model to model.

In Fig. \ref{fig7} we plot the proton over neutron density
fluctuations, normalized to their corresponding $\rho_p/\rho_n$
ratios for the largest unstable mode (first row), the
corresponding growth rates (second row) and the estimated size of
the clusters (third row) in $\beta$-equilibrium stellar matter
with trapped neutrinos for $T=5$, 10 and 12 MeV. 
These unstable  isoscalar modes, since protons and neutrons fluctuations are in phase,
are the ones that drive the system to a separation into a dense  symmetric phase and a neutron rich gas.
They  are different from the isovector modes represented in
Fig. \ref{fig2} at subsaturation densities,  for which neutron fluctuations are larger than proton fluctuations. This
property has already been point out in \cite{Greco-03}.


When the temperature is increased, the ratio of proton-neutron
density fluctuations decreases as well as the density range for
clusterization. On the other hand, there is an increase of the
average size of the clusters. From the top row of Fig. \ref{fig7}
we conclude that for warm  stellar matter with trapped neutrinos,
FSU and NL3$\omega\rho$ are less effective in recovering the
isospin symmetry in the liquid phase than the other two sets used.
This implies that within FSU and NL3$\omega\rho$  the  clusters are
neutron richer and the background gas has larger proton fractions,
when compared with TM1 and NL3. The presence of charged
particles in the gas will affect the electrical conductivity of
the crust. An overall conclusion from Fig. \ref{fig7} is that the
hotter the inner crust the larger and neutron richer are the
clusters and the smaller is the extension of the pasta phase.

It is interesting to notice the effect of temperature on the size
of the clusters: this behavior does not agree with the
conclusions of Ref. \cite{Avancini-08} which predicts a small
reduction of the cluster size with temperature (see Fig. 7 of
\cite{Avancini-08}). There, a simple approach was used to
determine the clusters: a zero surface thickness ansatz for the
clusters was used with a non self-consistent surface energy. It
is, therefore, important to understand this point by describing
the clusterized phase  within a  finite temperature
self-consistent Thomas-Fermi  calculation. In the approach used here
the increase of cluster sizes with temperature is a consequence of the
fact that the smaller wavelengths do not give rise to instabilities when temperature increases.
 At $T=12$ MeV only TM1 and FSU models
predict the formation of clusters. Globally, the effect of the
temperature is to reduce the instability region and increase the
cluster size.

Neutrino trapped matter is more isospin symmetric than neutrino free
$\beta$-equilibrium matter  and, therefore, the cluster size
estimates obtained from the different models do not differ  much
among them (see Fig. \ref{fig7} last row), quite different from
the estimates of cold $\beta$-equilibrium matter, Fig.
\ref{fig6}. This occurs because the isoscalar channel,
contrary to the isovector channel, of nuclear models at
subsaturation densities do not differ much.

\begin{figure}
   \begin{tabular}{c}
   \includegraphics[angle=0,width=0.8\linewidth]{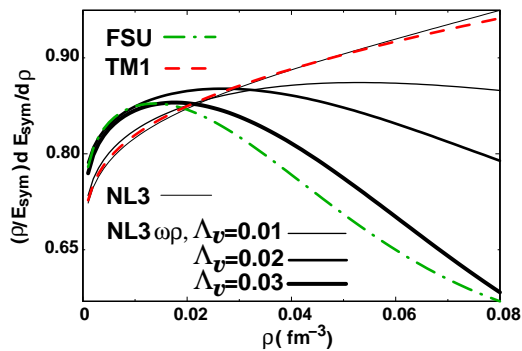}
   \end{tabular}
\caption{(Color online) The ratio $\frac{\rho}{\mathcal{E}_{sym}}
\frac{\mbox{d}\mathcal{E}_{sym}}{\mbox{d}\rho}$ 
 at subsaturation densities for the models under study.}
\label{fig8}
\end{figure}


The thermodynamical approach may clarify the behavior  shown in  Fig. \ref{fig7}a) for the
ratio of proton-neutron density fluctuations since the main conclusions taken
within both thermodynamic and dynamic  approaches are similar \cite{Ducoin-08,Xu-09-ASJ}.  In
\cite{Ducoin-08} it has been shown that the thermodynamic direction of instability  in first
order in the asymmetry $y=(\rho_n-\rho_p)/\rho$ is 
$$ \frac{\delta \rho_3}{\delta \rho_0}=\frac{\mathcal{E}_{sym}-\rho\,(\mbox{d}\mathcal{E}_{sym}/\mbox{d}\rho)
}{\mathcal{E}_{sym}- \rho {\cal F}^{''}_s/2}\,\, y,$$
where
${\cal F}^{''}$ is the second derivative of the free energy density of symmetric matter with respect to the density. Since the
isoscalar contribution of  the different models does not differ much at subsaturation densities
we may expect that the different behavior between the  models is determined by the
ratio   $\rho\,({\mbox{d}\mathcal{E}_{sym}}/{\mbox{d}\rho})/\mathcal{E}_{sym}$. We have plotted
this quantity  for the range of
densities of interest to understand the nonhomogenous phase. We conclude that 
{ this
quantity seems  also to define the behavior of the  ratio of the proton
to neutron density  fluctuations of the most unstable mode in a dynamical approach to the
instabilities of asymmetric nuclear matter.}


\section{Conclusions} \label{V}
In the present work we have compared normal modes properties described by
several relativistic nuclear models. We have chosen a set of
models with different characteristics, namely at large densities: NL3 has both a hard symmetry energy and a hard EOS;
TM1 has also a hard symmetry energy but a more soft EOS; for FSU, both the EOS and the symmetry energy
are soft and  NL3/TM1 with non-linear $\omega\rho$ terms have a hard/soft  EOS
at large densities and a symmetry energy which becomes softer when the coupling
$\Lambda_v$ increases.

Using the Vlasov equation we have determined the dispersion
relation of the collective modes of asymmetric neutral $npe$
matter. We have considered a)  collective modes at zero
temperature, both isovector and isoscalar-like, in a wide range of
densities and isospin asymmetries b) unstable modes at
subsaturation densities for both cold and warm matter. We have
analyzed the effect of the isovector properties, namely the
density dependence of the symmetry energy, on the collective
modes. All models have an isovector-like mode for densities
$\rho<0.2$ fm$^{-3}$ and $y_p\gtrsim 0.25$. However, when the
proton fraction decreases this mode becomes weaker and exists in a
smaller range of densities. Models with a smaller symmetry energy slope have a stronger mode with a larger sound
velocity. { It was shown that this was due to the larger coupling factor $f_\rho$ these models have at subsaturation densities.}

 For  $\rho> 0.2$ fm$^{-3}$, FSU has no collective mode.
The parametrization TM1 has a mode which keeps always its
isovector-like character, also for very high densities.
For NL3,  NL3$\omega\rho$ and TM1$\omega\rho$ the modes change  from
 isovector-like to isoscalar-like for a density above 0.2 fm$^{-3}$  that
depends both on the model and the proton fraction. At the crossing
the collective mode is a neutron wave. This feature was first
identified in \cite{Greco-03}. The differences among the models
are due to the different behaviors at large densities of their effective coupling contants $f_i$
and the nucleon effective mass $M^*$. 
A finite temperature calculation of these modes should be
performed because it may affect the evolution of a proto-neutron
star with trapped neutrinos.

For $\beta$-equilibrium cold neutrino free matter FSU has no
collective mode, TM1 has an isovector mode for all densities above
$\sim 0.11$ fm$^{-3}$, NL3 and NL3$\omega\rho$ have a collective
mode respectively above 0.15 and 0.3  fm$^{-3}$ which changes from
an isovector to an isoscalar-like mode at $\rho\sim 0.4$
fm$^{-3}$. It is seen that the models considered predict quite
different dynamical responses.

Next we have studied the unstable modes which all models present
at subsaturation densities, both for cold and warm matter. The
dynamical instability density region defines a lower bound for the
non-homogeneous matter in the crust of a compact star \cite{Oyamatsu-07,Xu-09-ASJ,Avancini-09}. We have
compared the previsions of the different models for cold
$\beta$-equilibrium matter and warm   $\beta$-equilibrium matter
with trapped neutrinos. It was shown that the non-homogeneous
phase could extend to $\sim 12$ MeV in some models but at $T=10$
MeV all models predicted still a pasta phase. While the extension
of cold $\beta$-equilibrium matter is model dependent and shows a
close correlation with the symmetry energy slope, for warm
$\beta$-equilibrium matter with trapped neutrinos all models
predict similar extensions, because of the small isospin
asymmetry. { However,  the set of models with the same isoscalar EOS, NL3 and NL3$\omega\rho$, show a decrease of the transition density with an increase of $L$ due to a corresponding increase of the isospin asymmetry at the transition.
 The systematic difference among all of them} is due to a different
distillation effect: models with a smaller slope of the symmetry
energy to symmetry energy ratio  predict a smaller distillation effect and, therefore, a proton richer
background gas. This will have an effect at
the level of transport properties such as electrical conductivity.
It was also shown that the average cluster size increases with
temperature. For cold  $\beta$-equilibrium matter the average
cluster size at the  crust-core transition density  does not depend on the
model and is 10 fm large. However, because the transition density
varies with the model there is quite a dispersion on the average
cluster size above 0.02 fm$^{-3}$: the models with the larger
pasta extension predict smaller clusters, $\sim 6$ fm, for a wider
range of densities. It was also shown that while the transition density decreases systematically with the increase of temperature, the transition pressure 
increases until $T\sim 5$ MeV and decreases for larger temperatures. No systematic behavior of the transition
pressure with the symmetry energy slope $L$ was obtained { if the whole set of models is considered both for neutrino-free matter and neutrino trapped matter. However, taking the set  of models 
with the same isoscalar EOS, NL3 and NL3$\omega\rho$, we could show that for neutrino-trapped matter
at $T=$0 MeV the pressure decreases when $L$ decreases due to the smaller isospin asymmetry, at $T=$ 5 MeV 
there was no systematic behaviour due to a competition between the lepton contribution and the nucleon EOS
 and at $T$ = 10 MeV the pressure decreases with an increase of $L$ due to the smaller lepton contribution, for 
the smaller densities.
}

We conclude that transport coefficients of stellar matter will be
sensitive to the density dependence of both the EOS of symmetric
matter and the density dependence of the symmetry energy and may
be used to set constraints on the EOS of cold and warm stellar
matter.

\appendix
\section{Equations for the fields}
\label{eq_campos}
\begin{equation}
\frac{\partial^2\phi}{\partial t^2} - \nabla^2\phi +m_s^2\phi +
\frac{\kappa}{2} \phi^2 + \frac{\lambda}{6} \phi^3
= g_s\rho_s({\bf r},t) \; ,
\label{eqmphi}
\end{equation}

\begin{eqnarray}
\frac{\partial^2 V_0}{\partial t^2} - \nabla^2 V_0 + m_v^2 V_0\, +\frac{1}{6}\xi g_v^4 V_0(V_\nu V^\nu)\,\nonumber \\
+2g_v^2g_\rho^2 \Lambda_{v} V_0(b_\nu b^\nu) =\, g_v j_0({\bf
r},t) \;, \label{eqmv0}
\end{eqnarray}
\begin{eqnarray}
\frac{\partial^2 V_i}{\partial t^2} - \nabla^2 V_i + m_v^2 V_i\, +\frac{1}{6}\xi g_v^4 V_i(V_\nu V^\nu)\,\nonumber \\
+2g_v^2g_\rho^2 \Lambda_{v} V_i(b_\nu b^\nu) =\, g_v j_i({\bf
r},t) \;, \label{eqmv}
\end{eqnarray}
\begin{equation}
\frac{\partial^2 b_0}{\partial t^2} - \nabla^2 b_0 + m_\rho^2
b_0\,+2g_v^2g_\rho^2 \Lambda_{v} b_0(V_\nu V^\nu) =\,
\frac{g_\rho}{2} j_{3,0}({\bf r},t)\;, \label{eqmb0}
\end{equation}
\begin{equation}
\frac{\partial^2 b_i}{\partial t^2} - \nabla^2 b_i + m_\rho^2
b_i\,+2g_v^2g_\rho^2 \Lambda_{v} b_i(V_\nu V^\nu) =\,
\frac{g_\rho}{2} j_{3,i}({\bf r},t) \;, \label{eqmb}
\end{equation}
\begin{equation}
\frac{\partial^2 A_0}{\partial t^2} - \nabla^2 A_0 \, =\,
e [j_{0p}({\bf r},t) -j_{0e}({\bf r},t)] \;,
\label{eqma0}
\end{equation}
\begin{equation}
\frac{\partial^2 A_i}{\partial t^2} - \nabla^2 A_i \, =\,
e[j_{ip}({\bf r},t) -j_{ie}({\bf r},t) ]\;,
\label{eqma}
\end{equation}
where the scalar density is
\begin{eqnarray*}
\rho_s({\bf r},t)&=&2\sum_{i=p,n}\int\frac{d^3p}{(2\pi)^3}\,
(f_{i+}({\bf r},{\bf p},t)+f_{i-}({\bf r},{\bf p},t))\,
\frac{M_i^*}{\epsilon_i} \\ \nonumber
&=&\rho_{sp}+\rho_{sn}
\end{eqnarray*}
and the isovector density is
\begin{eqnarray*}
\rho_{3s}({\bf r},t)&=&2\sum_{i=p,n}\int\frac{d^3p}{(2\pi)^3}\,\tau_i
(f_{i+}({\bf r},{\bf p},t)+f_{i-}({\bf r},{\bf p},t))\,
\frac{M_i^*}{\epsilon_i} \qquad \\ \nonumber
&=&\rho_{sp}-\rho_{sn}.
\end{eqnarray*}
The components of the baryonic four-current density are
\begin{eqnarray*}
j_0({\bf r},t)&=&2 \sum_{i=p,n}\int\frac{d^3p}{(2\pi)^3}\,
(f_{i+}({\bf r},{\bf p},t)-f_{i-}({\bf r},{\bf p},t))\\ \nonumber
&=& \rho_p+\rho_n \;,
\end{eqnarray*}
\begin{equation*}
{\bf j}({\bf r},t)=2\sum_{i=p,n}\int\frac{d^3p}{(2\pi)^3}\,
(f_i({\bf r},{\bf p},t)+f_{i-}({\bf r},{\bf p},t))
\frac{{\bf p}-\boldsymbol{\cal V}_i}{\epsilon_i} \;,
\end{equation*}
\begin{equation*}
j_{0e}({\bf r},t)=2 \int\frac{d^3p}{(2\pi)^3}\,
(f_{e+}({\bf r},{\bf p},t)-f_{e-}({\bf r},{\bf p},t))\; ,
\end{equation*}
\begin{equation*}
{\bf j}_e({\bf r},t)=2\int\frac{d^3p}{(2\pi)^3}\,
(f_{e+}({\bf r},{\bf p},t)+f_{e-}({\bf r},{\bf p},t)\,
\frac{{\bf p}+e\boldsymbol{\mathbf A}}{\epsilon_e} \; ,
\end{equation*}
and the components of the isovector four-current density are
\begin{eqnarray*}
j_{3,0}({\bf r},t)&=&2\sum_{i=p,n}\int\frac{d^3p}{(2\pi)^3}\, \tau_i
(f_{i+}({\bf r},{\bf p},t)-f_{i-}({\bf r},{\bf p},t))\\ \nonumber
&=&\rho_p-\rho_n \;, \\ \nonumber
{\bf j}_3({\bf r},t)&=&2\sum_{i=p,n}\int\frac{d^3p}{(2\pi)^3}\,
\frac{{\bf p}-\boldsymbol{\cal V}_i}{\epsilon_i}\\ \nonumber
&\times& \tau_i\, (f_{i+}({\bf r},{\bf p},t)+f_{i-}({\bf r},{\bf p},t)) \; ,
\end{eqnarray*}
 with
$\epsilon_i=\sqrt{({\bf p}-\boldsymbol{\cal V}_i)^2+{M_i^*}^2} \;, i=p,n \quad$
and \\
$\epsilon_e=\sqrt{({\bf p}+e{\mathbf A})^2+m_e^2} \;.$
\section{Equations of motion}
\label{eq_flut}
Defining
$$
\delta \rho_{si}=\int \frac{d^3p}{(2 \pi)^3}\frac{p \cos \theta}
{\epsilon_{0 i}} \left[S_{\omega +}^i \frac{df_{0i+}}{dp^2} +
S_{\omega -}^i \frac{df_{0i-}}{dp^2} \right],
$$
$$
\delta \rho_i=\int \frac{d^3p}{(2 \pi)^3} p \cos \theta
\left[S_{\omega +}^i \frac{df_{0i+}}{dp^2} -
S_{\omega -}^i \frac{df_{0i-}}{dp^2} \right]
$$
with
$$ \frac{df_{0i \pm}}{dp^2} = \frac{1}{2T \epsilon_{0 i}} f_{0i \pm}
(f_{0i \pm} -1)$$
the equations of motion for the field fluctuations read
\begin{eqnarray}
\left( \omega^2-k^2-m^2_{s,eff}\right) \delta\phi_\omega &=&
4i g_s k \sum_{i=p,n} M_i^* \delta \rho_{si}  \label{fi} \\
\left( \omega^2-k^2-m_{v,eff}^2 \right)\delta V_\omega^0 &=&4g_v^2g_\rho^2\Lambda_{v} V_0^{(0)} b_0^{(0)}\delta b_\omega^0 \nonumber \\
 &+&4i g_v k \sum_{i=p,n} \delta \rho_i \label{V0} ,\\
\left( \omega^2-k^2-m_{\rho,eff}^2 \right)\delta b_\omega^0 &=&4g_v^2g_\rho^2\Lambda_{v} V_0^{(0)} b_0^{(0)}\delta V_\omega^0 \nonumber \\
&+&2i g_\rho k \sum_{i=p,n} \tau_i \delta \rho_i \label{b0} ,\\
\left( \omega^2-k^2 \right)\delta A_\omega^0 &=&
4i e k \sum_{i=p,e} (-1)^{n_i}\delta \rho_i,
\label{A}
\end{eqnarray}
and from the continuity equation for the density currents, we get for the components of the vector fields
\begin{eqnarray}
\omega\, \delta V_\omega^0 &=& k\, \delta V_\omega \label{contV} ,\\
\omega\, \delta b_\omega^0 &=& k\, \delta b_\omega \label{contb} ,\\
\omega\, \delta A_\omega^0 &=& k\, \delta A_\omega\,.
\label{contA}
\end{eqnarray}

\section{Solutions for the eigenmodes and the dispersion relation}
\label{rel_disp}

We define the following quantities
$$\bar \omega=\frac{\omega}{k}, \quad x= \frac{p \cos \theta}{\epsilon_{0 i}},\quad \\ G_{\phi}=\frac{g_s M^*}{k}, \quad G_{\nu}=\frac{g_v}{k}, \quad G_{\rho_i}=\frac{g_\rho \tau_i}{2k},$$
$$Z_{\omega\rho}=\frac{1}{\left[(\bar\omega^2-\bar\omega_\rho^2)(\bar\omega^2-\bar\omega_v^2)-D_{\omega\rho}^2\right]},$$
$$D_{\omega\rho}=\frac{4 g_v^2 g_\rho^2 \Lambda_{v} V_0^{(0)} b_0^{(0)}}{k^2},$$
$$c_s=\frac{G_\phi^2}{2 \pi^2 T(\bar\omega^2-\bar\omega_s^2)} ,\quad  \bar\omega_s^2=\frac{1}{k^2}(k^2+m^2_{s,eff}) $$

\begin{eqnarray*}
c_\omega^{ij}&=&\frac{Z_{\omega\rho}}{2 \pi^2 T}(1-\bar\omega^2)\left[G_\nu\left((\bar\omega^2-\bar\omega_\rho^2)G_\nu+D_{\omega\rho} G_{\rho_j}\right)\right.\\
&+&\left. G_{\rho_i}\left((\bar\omega^2-\bar\omega_v^2)G_{\rho_j}+D_{\omega\rho} G_\nu\right)\right]
\end{eqnarray*}

$$\bar\omega_v^2=\frac{1}{k^2}(k^2 + m_{v,eff}^2), \quad \bar\omega_{\rho}^2=\frac{1}{k^2}(k^2+m_{\rho,eff}^2) $$

$$c_e= \frac{-2}{(2 \pi)^2 T} \left(\frac{e}{k}\right)^2,$$
$$I_{\omega_\mp}(\epsilon)= \int_{-p/\epsilon}^{p/\epsilon} dx
\frac{x}{\bar \omega \pm x} = \pm \left[2 \frac{p}{\epsilon} + \bar \omega ~ln
\left| \frac{\bar \omega - p/\epsilon}{\bar \omega + p/\epsilon} \right|
\right],$$
$$I^{ni}_{\omega_\mp}= \int_{M_i^*}^{\infty} \epsilon^n I_{\omega_\mp}(\epsilon)
f_{0 i \mp}( f_{0 i \mp} -1) d\epsilon\, ,$$
$$A_{\omega i,\pm}^n=\int_{M_i^*}^\infty \epsilon^nd\epsilon
\int_{-p/\epsilon}^{p/\epsilon}
dx\, x\, S_{\omega \pm}^i(x,p)\, f_{0i,\pm}\left( f_{0i,\pm}-1 \right),$$
\begin{eqnarray}
\rho_{\omega i}&=&A_{\omega i +}^1-A_{\omega i -}^1, \quad i=p,n,e \label{rhoi}\\
\rho^S_{\omega i}&=&A_{\omega i +}^0+A_{\omega i -}^0, \quad i=p,n\, . \label{rhosi}
\end{eqnarray}

The coefficients $a_{ij}$ are defined as:
\begin{eqnarray*}
a_{11}&=&1+c_s\left( I_{\omega  +}^{0p}- I_{\omega  -}^{0p} \right),\,\,\,
a_{12}=c_s\left(  I_{\omega  +}^{0p}- I_{\omega  -}^{0p} \right),\\
a_{13}&=&-(c_\omega^{pp}+c_e) \left(I_{\omega  +}^{1p}+I_{\omega-}^{1p}\right),\\
a_{14}&=&-c_\omega^{pn} \left(I_{\omega  +}^{1p}+I_{\omega -}^{1p}\right),\,\,
a_{15}= c_e \left(I_{\omega  +}^{1p}+I_{\omega -}^{1p}\right),\\
a_{21}&=&c_s\left(  I_{\omega  +}^{0n}- I_{\omega  -}^{0n} \right),\,\,
a_{22}=1+c_s\left(  I_{\omega  +}^{0n}- I_{\omega  -}^{0n} \right),\\
a_{23}&=&-c_\omega^{np} \left(I_{\omega  +}^{1n}+I_{\omega-}^{1n}\right),\\
a_{24}&=&-c_\omega^{nn} \left(I_{\omega  +}^{1n}+I_{\omega -}^{1n}\right),\,\,
a_{25}=0,\\
a_{31}&=&c_s\left( I_{\omega+}^{1p}+ I_{\omega  -}^{1p} \right),\,\,
a_{32}=c_s\left( I_{\omega+}^{1p}+ I_{\omega  -}^{1p} \right),\\
a_{33}&=&1-(c_\omega^{pp}+c_e) \left(I_{\omega +}^{2p}-I_{\omega-}^{2p}\right),\\
a_{34} &=&-c_\omega^{pn} \left(I_{\omega  +}^{2p}-I_{\omega -}^{2p}\right),\,\,
a_{35}=c_e \left(I_{\omega  +}^{2p}-I_{\omega -}^{2p}\right),\\
a_{41}&=&c_s\left(  I_{\omega  +}^{1n}+ I_{\omega  -}^{1n} \right),\,\,
a_{42}=c_s\left(  I_{\omega  +}^{1n}+ I_{\omega  -}^{1n} \right),\\
a_{43}&=&-c_\omega^{np} \left(I_{\omega  +}^{2n}-I_{\omega-}^{2n}\right),\\
a_{44}&=&1-c_\omega^{nn} \left(I_{\omega  +}^{2n}-I_{\omega -}^{2n}\right),\,\, a_{45}=0,\\
a_{51}&=&a_{52}=a_{54}=0,\\
a_{53}&=&c_e \left(I_{\omega  +}^{2e}-I_{\omega-}^{2e}\right),\,\,
a_{55}=1-c_e \left(I_{\omega  +}^{2e}-I_{\omega -}^{2e}\right).
\end{eqnarray*}

The ratios of the amplitudes are given by:
\begin{equation}
\frac{\rho_{\omega p}}{\rho_{\omega n}}=-\frac{a_{11}a_{pn}+a_{12}a_{nn}+a_{14}}{a_{11}a_{pp}+a_{12}a_{np}+a_{13}-a_{15}a_{53}/a_{55}} \label{amp1}
\end{equation}
with
\begin{equation*}a_{pp}=\frac{a_{22}a_{43}-a_{23}a_{42}}{a_{21}a_{42}-a_{22}a_{41}},\, \quad a_{pn}=\frac{a_{22}a_{44}-a_{24}a_{42}}{a_{21}a_{42}-a_{22}a_{41}},\,\end{equation*}

\begin{equation*}
a_{nn}=\frac{a_{44}a_{21}-a_{41}a_{24}}{a_{41}a_{22}-a_{42}a_{21}},\,  \quad
a_{np}=\frac{a_{43}a_{21}-a_{41}a_{23}}{a_{41}a_{22}-a_{42}a_{21}}
\end{equation*}
and
\begin{equation}
\frac{\rho_{\omega e}}{\rho_{\omega p}}=-\frac{a_{53}}{a_{55}} \label{amp2}.
\end{equation}

\section*{ACKNOWLEDGMENTS}

This work was partially supported by FCT (Portugal) under grants PTDC/FIS/64707/2006, SFRH/BPD/29057/2006 
and CERN/FP/109316/2009, and by COMPSTAR, an ESF Research Networking Programme.

\end{document}